\documentclass[preprint,12pt]{elsarticle}

\usepackage{amssymb}
\usepackage{amsmath,amssymb,amsfonts}
\usepackage{xcolor}
\usepackage{ulem}
\usepackage{setspace}
\usepackage{cancel}
\usepackage{subfig}
\usepackage{array} 
\usepackage{verbatim}
\usepackage{graphicx}
\usepackage{longtable,booktabs}
\usepackage{bbding}
\usepackage{utfsym}
\usepackage{setspace}
\usepackage{algorithm}
\usepackage{tabularx}
\usepackage{adjustbox}
\usepackage[noend]{algpseudocode}

\algrenewcommand\algorithmiccomment[1]{\hfill\(\#\)\;#1}
\usepackage{array,booktabs,tabularx,makecell,adjustbox,caption}
\usepackage{ragged2e}
\newcolumntype{L}[1]{>{\raggedright\arraybackslash}p{#1}}

\makeatletter
\def\ps@pprintTitle{%
 \let\@oddhead\@empty
 \let\@evenhead\@empty
 \let\@oddfoot\@empty
 \let\@evenfoot\@empty}
\makeatother
\begin{document}
\captionsetup[figure]{labelfont={bf},labelformat={default},labelsep=period,name={Fig.}}
\begin{frontmatter}



\title{Joint Attention Mechanism Learning to Facilitate Opto-physiological Monitoring during Physical Activity}


\affiliation[inst1]{organization={Wolfson School of Mechanical, Electrical and Manufacturing Engineering},
            addressline={Loughborough University}, 
            city={Loughborough},
            postcode={LE11 3TU}, 
            country={UK}}
            
\affiliation[inst2]{organization={School of Sport, Exercise and Health Sciences},
            addressline={Loughborough University}, 
            city={Loughborough},
            postcode={LE11 3TU}, 
            country={UK}}
            


\author[inst1]{Xiaoyu Zheng}
\author[inst1]{Sijung Hu\corref{mycorrespondingauthor}}
\cortext[mycorrespondingauthor]{Corresponding author}
\ead{s.hu@lboro.ac.uk}
\author[inst1]{Vincent Dwyer}
\author[inst2]{Laura Barrett}
\author[inst1]{Mahsa Derakhshani} 

\begin{abstract}

Opto-physiological monitoring including photoplethysmography (PPG) provides non-invasive cardiac and respiratory measurements, yet motion artefacts (MAs) during physical activity degrade its signal quality and downstream estimation concurrently. An attention-mechanism-based generative adversarial network (AM-GAN) was proposed to model motion artefacts and mitigate their impact on raw PPG signals. The AM-GAN learns how to transform motion-affected PPG into artefact-reduced waveforms to align with triaxial acceleration signals corresponding to artefact components gained from a triaxial accelerometer. The AM-GAN has been validated across four experimental protocols with 43 participants performing activities from low to high intensity (6--12km/h). With the public datasets, the AM-GAN achieves mean absolute error (MAE) for heart rate (HR) of 1.81 beats/min on IEEE-SPC and 3.86 beats/min on PPGDalia. On the in-house LU dataset, it shows the MAEs \textless 1.37 beats/min for HR and 2.49 breaths/min for respiratory rate (RR). A further in-house C2 dataset with three oxygen levels (16\%, 18\%, and 21\%) was applied in the AM-GAN  to attain a MAE of 1.65\% for SpO$_2$. The outcome demonstrates that the AM-GAN offers a robust and reliable physiological estimation under various intensities of physical activity.
\end{abstract}




\end{frontmatter}


\section{Introduction}
\label{sec:intro}
The pulsatile waveform extracted from the PPG signal is a key component in obtaining physiological parameters, including heart rate (HR), respiration rate (RR) and oxygen saturation (SpO$_2$). Obtaining reliable physiological parameters from PPG relies upon a high-quality signal.  Motion artefacts (MAs), frequently generated during physical activity, impede the acquisition of clear PPG signals, thereby diminishing the stability and accuracy of vital sign measurements. MAs can arise from changes in sensor, skin contact due to movement, as well as from haemodynamic effects \cite{b3turcott2008hemodynamic}. Various techniques have been investigated to extract MA to produce MA-reduced PPG signals. These methods include Empirical Mode Decomposition (EMD), Fast Fourier Transform (FFT), Independent Component Analysis (ICA), wavelet denoising \cite{b4pankaj2022review}, Adaptive Noise Cancellation (ANC) with Recursive Least Squares (RLS), Adaptive-Size Least Mean Squares (AS-LMS) \cite{b5ram2011novel} and an efficient envelope-based PPG denoising algorithm (EPDA) \cite{b36bradley2024opening}. These approaches tend to underperform during medium-to-high-intensity physical activities, a challenge addressed by the TROIKA algorithm \cite{b6zhang2014troika}. Also, such MA-removal methods are complex and a challenge for efficient embedding in a wearable system \cite{b7nabavi2020robust}. Furthermore, a lightweight motion artefact intensity classification and removal (MAICR) framework is proposed to extract HR using acceleration time-domain information and PPG frequency-domain information \cite{b34hao2024ppg}. Zhang et al. \cite{b35zhang2024removal} proposed an MA-removal method combining CEEMDAN-based multiscale permutation entropy (CEEMDAN-MPE) with variable-step-size least mean squares (VS-LMS), effectively enhancing the accuracy of SpO$_2$ measurement. Although these studies adopt an acceleration reference for MA removal, the acceleration reference usually fails to precisely capture MA associated with subtle finger or wrist movements and gestures, leading to potentially uncorrelated MA behaviour. There has been a growing demand for additional, reliable physiological parameters beyond HR estimates, such as RR and SpO$_2$, where obtaining a higher-quality pulsatile waveform is key. Consequently, additional PPG algorithms are required to extract these additional parameters/biomarkers during physical activities that are advantageous during wearable opto-physiological monitoring. Such a prospect may well lie in the domain of Neural Networks. 

Inspired by the rapid development of deep-learning (DL) techniques, several DL algorithms to extract physiological parameters from raw PPG signals have been recently presented. These techniques can be categorised into two types: \textit{end-to-end} methods and \textit{feature extraction} methods. The former directly establishes a mapping from raw PPG signals to the target physiological parameters \cite{b8luque2018end}, while the latter obtains the target parameters by extracting features pre-processed from raw PPG signals \cite{b9biswas2019cornet}. Data-driven, with strong fitting abilities, these neural networks often outperform traditional methods \cite{b10esgalhado2021application}. However, most DL end-to-end based methods estimate specific physiological parameters directly from raw PPG, while neglecting the reconstruction of high-quality PPG waveforms that underpin reliable parameter estimation.

Some recent studies have demonstrated that using DL methods can benefit the noise and motion detection from the input noisy PPG signals \cite{b11goh2020robust}. Specifically, a one-dimensional (1D) convolutional neural network (CNN) has been used to learn intrinsic PPG features and classify signals as clean or artefact-contaminated \cite{b11goh2020robust}. Other DL-based approaches convert PPG signals into two-dimensional (2D) representations and use 2D CNNs for signal-quality classification at rest and during low-intensity exercise. \cite{b12chen2021signal}, \cite{b13afandizadeh2023accurate}. Although these approaches have demonstrated high accuracy in HR evaluation, their effectiveness during medium-to-high-intensity activities has not yet been assessed. Evaluating performance in such scenarios is crucial, as MAs during these activities are more complex and exhibit diverse morphologies. Currently, attention mechanisms have become central to modern sequence modelling and sequence-to-sequence tasks, enabling models to capture long-range dependencies regardless of position. \cite{b30ashish2017attention}. Panagiotis et al. \cite{b37kasnesis2023feature} proposed PULSE, a deep learning method leveraging temporal convolutions and multi-head attention to enhance HR monitoring. Additionally, Christodoulos et al. \cite{b38kechris2024kid} introduced the adaptive Q-PPG \cite{b31burrello2021q} framework, combining the baseline Q-PPG model with an attention mechanism to improve HR tracking accuracy. Recent work combining generative adversarial networks (GANs) with attention has shown promise. For example, \cite{b31sarkar2021cardiogan} uses an attention-based generator to identify local salient features and employs dual discriminators to preserve fidelity in both the time and frequency domains when synthesising ECG from PPG. A general limitation of \cite{b31sarkar2021cardiogan} is that MAs are not considered. In contrast, a GAN was employed to remove low-intensity motion artefacts from PPG signals by converting the signals into two-dimensional correlation images, enabling the use of a standard image-based GAN architecture \cite{b13afandizadeh2023accurate}. Similarly, the elimination of ocular artefacts from electroencephalography (EEG) data, crucial for numerous brain-computer interface applications, was explored in \cite{b17sawangjai2021eeganet}. Furthermore, Sawangjai et al. \cite{b33sawangjai2025removal} proposed an attention GAN with dual discriminator to remove MAs without additional motion data from accelerometers or gyroscopes. However, their approach faces an obvious limitation in effectively addressing MAs during medium-to-high-intensity movements due to inherent architectural constraints.

To tackle these challenges, the study proposes an attention mechanism-based learning approach capable of robustly extracting physiological parameters from multi-sensor data, specifically PPG and 3-axis acceleration signals, across multiple datasets.  Hence, an attention mechanism combined with a GAN is chosen and utilised for multi-sensor fusion to effectively remove MAs from raw PPG signals during physical activities ranging from low to high intensity. Pre-filtered, motion-contaminated PPG signals from an optoelectronic patch sensor, triaxial acceleration signals from an accelerometer sensor, and associated velocities were adopted as the inputs of the proposed AM-GAN model. Once the model learns how to map the contaminated PPG signals to MA-free PPG signals, the generator generates a PPG signal similar to the ground-truth PPG signal. At the same time, the discriminator distinguishes the generated PPG signal from the ground-truth PPG signal. The attention mechanism is well-suited for multi-sensor fusion, significantly enhancing the generator's capability to accurately identify how MAs the characteristics of PPG waveforms. Consequently, this improves the system's effectiveness in removing MAs in physical activities ranging from low to high intensity. The relevant literatures are summarised in Appendix \ref{tab:lit_review}. The contributions of this work are outlined as follows:

\begin{enumerate}
\item  We propose an updated attention mechanism integrated within a GAN specifically designed for multi-sensor fusion, enabling the effective removal of both in-band and out-of-band noise from PPG signals.

\item  The capability of the attention-based GAN (AM-GAN) to generate clear and reliable PPG waveforms by removing MAs from raw signals using synchronised triaxial accelerometer data across various physical activities, ranging from low to high intensity within a frequency range of 0.1–4.0 Hz is evaluated.

\item  The effectiveness of the proposed AM-GAN model is validated through the extraction of HR, RR, and SpO$_2$. Comprehensive performance comparisons against state-of-the-art methods are conducted using both intra-dataset and cross-dataset evaluations.

\item  The practical perspective is demonstrated while employing the attention mechanism by comparing outputs in both time and frequency domains, highlighting improvements achieved through the proposed approach.
\end{enumerate}

\section{Methods and Materials}
\begin{figure}[!hbt]
    \centering
    \includegraphics[width=0.8\textwidth]{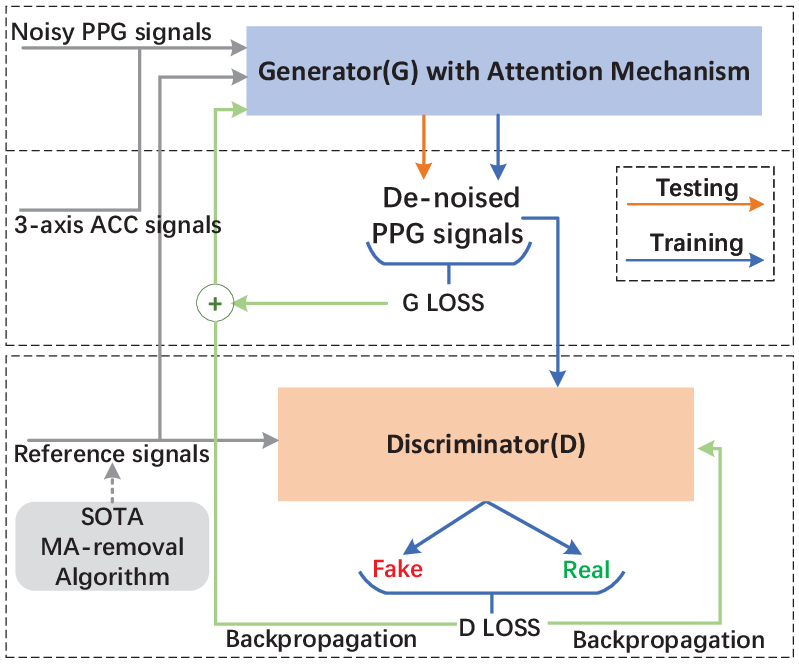}
    \caption{The proposed AM-GAN comprising a joint attention mechanism together with a GAN.}
    \label{Fig1}
\end{figure}
In this study, we propose an AM-GAN (depicted in Fig.~\ref{Fig1}) comprising an attention mechanism together with a GAN to learn the mapping from the MA-corrupted PPG signal to the MA-free PPG signal. Particularly, the proposed AM-GAN consists of three main parts: 1) Generator ($G$) together with the Attention Mechanism, 2) Discriminator ($D$), and 3) Motion removal algorithm (during training). 

During the training, the generator takes an MA-corrupted PPG signal $p(t)\in\mathcal{P}_{Tr}$ (where $\mathcal{P}_{Tr}$ denotes the training data set) as an input as well as triaxial acceleration signals $a_{x,y,z}(t)$ as a conditional input and generates an approximation to the MA-free PPG signal $s(t)=G(p(t), a_{x,y,z}(t))$ as its output. An attention mechanism guides the generator to focus on the most salient features in the encoded PPG signal. The discriminator is designed to distinguish the ground truth $s_{ref}(t)$ from the signals $s(t)=G(p(t), a_{x,y,z}(t))$. Finally, the motion removal algorithm, MR algorithm \cite{b21zheng2023rapid}, extracts the reference PPG signals from the MA-corrupted PPG signal $p(t)$, i.e., $s_{ref}(t)=MR(p(t))$, for training purposes.

To verify the performance of the trained network, the output of the generator is compared to the reference signal for signals in the testing data set, i.e., $p(t)\in\mathcal{P}_{Te}$. The physiological parameters (HR, RR, and SpO$_2$) are also extracted from the MA-free PPG signals and compared with the ones from reference signals. The details of the proposed AM-GAN are described in the following subsections.

\subsection{U-Net Generator Aligned with Attention Mechanism}
\begin{figure*}[!hbt]
    \centering
    \includegraphics[width=1\textwidth]{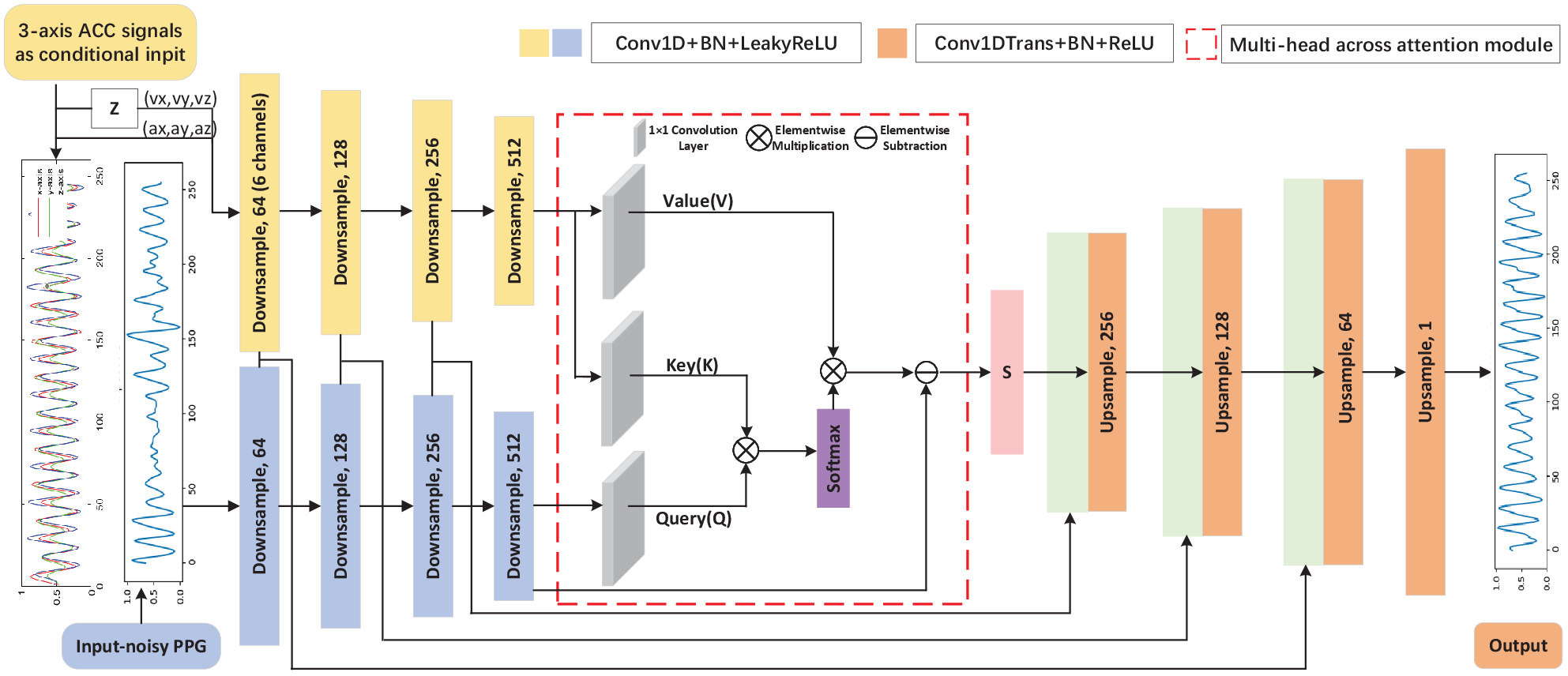}
    \caption{The generator network aligned with the attention mechanism. In the cross-attention mechanism, the noisy PPG (P) feature map is set as the \textit{Query} tensor, while the combined feature maps for the triaxial ACC and VEL are set as \textit{Key} and \textit{Value} tensors.}
    \label{Fig2}
\end{figure*}

The U-Net generator was initially proposed for image classification \cite{b20ronneberger2015u}. Fig. \ref{Fig2} shows the modified U-Net generator as a de-noising encoder-decoder with skip connections and the attention mechanism to take the noisy PPG signal as an input and the triaxial acceleration (ACC) as the conditional input. Meanwhile, the triaxial velocity (VEL) is created using the acceleration signal and constant matrix $Z$, i.e., $v_x=\frac{1}{f_s}Z[a_x]$, where $f_s$ is the sampling rate and $Z$ is the lower triangular matrix of constants (constant is 1). 

In the modified U-Net generator architecture, 1D convolution and deconvolution functions are utilised instead of the typical 2D functions, and the latent vector $z$ is omitted, or more properly is replaced in essence by the motion reference proxy obtained from the triaxial ACC and VEL. These changes adapt the network better for processing 1D signals and help to minimise overfitting during the generation of the targeted PPG signals. Specifically, the encoding phases include four convolutional layers, each followed by batch normalisation (BN) and leaky rectified linear units (LeakyReLU). These components are integrated within a strided convolution process using a stride of two. The decoding process is the reversal of the encoding process, which is implemented through deconvolution layers followed by batch normalisation (BN) and rectified linear (ReLU) units to decode the de-noised PPG feature map culminating in the acquisition of the desired de-noised PPG signal. Additionally, the generator network employs skip connections that link each encoding layer to its corresponding decoding layer, effectively circumventing the central compression stage of the generator network. These skip connections enable the direct transfer of fine-grained features from the encoding stage to the decoding stage, addressing the potential loss of low-level details that can occur when all feature information passes through the central compression bottleneck of the network\cite{b20ronneberger2015u}. In the AM-GAN generator structure, the PPG signal encoding layer and motion reference encoding layer are concatenated to link the corresponding decoding layer. 

Instead of directly merging features calculated in the contracting path at the same hierarchical level within the upsample block, the attention mechanism is employed for multi-sensor fusion to identify the MAs and target the relevant PPG features from the downsample block. This could be achieved by using the conditional motion reference signal, i.e., encoded triaxial ACC with related VEL, to guide the attention over the encoded input-noisy PPG signal (P).

\subsection{Attention Mechanism}
\subsubsection{Problem Formulation}
Referring to the simplest version of our MR model~\cite{b21zheng2023rapid}, the noise $n[i]$, in the measured noisy PPG signal $p[i]$, can be approximated by a linear combination of acceleration outputs $a_x[i], a_y[i], a_z[i]$ and associated velocity (the integrated outputs $v_x[i], v_y[i]$ and $v_z[i]$) from the triaxial ACC, thus $n[i] =\boldsymbol{a}[i]\Xi$  where the columns of the $(N\times 6)$ array $\boldsymbol{a}[i]$are the six motion signals with coefficients given by the $(6\times 1)$ array $\boldsymbol{\Xi}$. An orthonormal basis for this subspace may be obtained from the six columns of the $(N \times 6)$ array $\boldsymbol{\Phi}=\boldsymbol{a}\boldsymbol{U}^\intercal\boldsymbol{D^{-1/2}}$, where $\boldsymbol{U}$ and $\boldsymbol{D}$ are the unitary matrix that diagonalises the $6\times 6$ array $\boldsymbol{a}^\intercal\boldsymbol{a}$, and the resulting diagonal form (i.e. $\boldsymbol{a}^\intercal\boldsymbol{a}=\boldsymbol{U}^\intercal\boldsymbol{D}\boldsymbol{U}$). The motion-suppressed PPG signal is then rewritten as the more obvious: 
\begin{equation}
\label{MinRes}
    \boldsymbol{s}= \boldsymbol{p} - \boldsymbol{\Phi}\boldsymbol{\Phi}^\intercal \boldsymbol{p}
\end{equation}
where $\boldsymbol{\Phi}$ plays the role of a unit vector in standard Euclidean geometry.  From \eqref{MinRes}, $\boldsymbol{s}^\intercal\boldsymbol{n}=0$, meaning that the method assumes that the clean signal and the noise are separable in the frequency domain. In the case that the spectra of $\boldsymbol{s}$ and $\boldsymbol{n}$ do overlap, \eqref{MinRes} should remove some, but not necessarily all, of the MAs; it will represent motion suppression rather than motion removal. 

One advantage of choosing this method to trial the hypothesis that 'Motion Removal' algorithms may be learned by an NN is its relative insensitivity to data scaling, which occurs in standard neural network pre-processing, often to improve convergence. This maintains a focus on the behaviour of the network to mimic the algorithm. This is clear from the fact that the orthonormal basis $\boldsymbol{\Phi}$ in \eqref{MinRes} is independent of any individual scaling of the motion components because the scaled accelerometer reading $\boldsymbol{a}'=\boldsymbol{a}\boldsymbol{\Lambda}$, for any diagonal $(6 \times 6)$ scaling matrix $\boldsymbol{\Lambda}$, occupies the same subspace as $\boldsymbol{a}$ and so has $\boldsymbol{\Phi}$ as a basis (more precisely the array $\boldsymbol{\Phi}\boldsymbol{\Phi}^\intercal$ is unchanged). Likewise, \eqref{MinRes} is linear in $\boldsymbol{p}$ so that scaling $\boldsymbol{p}'=\lambda \boldsymbol{p}$ simply scales $\boldsymbol{s}$ by $\lambda$ also. Second, the final term in \eqref{MinRes} demonstrates a simple mathematical form of ‘traditional attention mechanism’.

\subsubsection{Cross-attention Mechanism for Multi-sensor Fusion}
A self-attention mechanism was proposed as a core component of the transformer model \cite{b30ashish2017attention}. The self-attention accepts a tensor comprising sequential data and establishes its self-correlation. Specifically, it involves linear projections of the input $X$ into  \textit{queries} ($Q$),  \textit{keys} ($K$), and  \textit{values} ($V$). The scaled dot-product attention ($A$) is then applied to these projections as $A = softmax((Q K^T) / \sqrt{d_k}) V$, here $d{_k}$ is the dimensionality of the queries and keys.


In contrast to the self-attention mechanism that correlates the input $X$ with itself, the multi-head cross-attention mechanism is proposed for multi-sensor fusion. They are two inputs $X_1$ and $X_2$, where $X_1$ represents \textit{query} $Q$ tensors and $X_2$ represents the \textit{key} $K$ and \textit{value} $V$ tensors. In the proposed attention mechanism, these tensors are related to different sensor inputs, including encoded noisy PPG and encoded triaxial ACC with related VEL feature maps after four downsample blocks of the generator. The encoded feature maps are fed into the attention mechanism, as shown in Fig. \ref{Fig2} and Algorithm \ref{algo1}. Algorithm \ref{algo1} presents the pseudocode implementation of our proposed multi‑sensor fusion cross‑attention mechanism. Specifically, the proposed attention mechanism takes the generated PPG encoded feature map processed by a (feature pooling) $1\times1$ convolution layer to create the \textit{query} $Q$. Also, the \textit{key} $K$ and \textit{value} $V$ pair is formed through a similar $1\times1$ convolution layer of generated combined encoded feature map with triaxial ACC and the related VEL. The $K$ and $V$ vectors are used to calculate the attention weights, indicating to the model which parts of the MAs to focus on. The $Q$ vector is the information that these attention weights will be applied to, with the aim of highlighting the MA features in the PPG encoded feature map that are most relevant to the ACC and VEL combined encoded feature map. 

Inspired by the previous successful MR model and the structure of\eqref{MinRes}, the modified multi-head cross-attention mechanism is utilised as an alternative to the multi-sensor fusion and the traditional computation of $\boldsymbol{\Phi}$ matrix for focusing on the MAs at different exercise intensities. Furthermore, the target output of the modified cross-attention mechanism $S$ is calculated by:
\begin{equation}
\label{e4}
S = P \ominus softmax((QK^T) / \sqrt{d_k}) V
\end{equation}
where $\ominus$ represents the elementwise subtraction. Specifically, the inner product of $Q$ and $K$ depends upon the angle between the two vectors. In other words, it is the projection of one vector on the other vector, which means that the larger the projection value, the higher the correlation between the two vectors. As delineated in \eqref{e4}, the process involves computing the correlation between $Q$ (encoded PPG feature maps) and $K$ (encoded motion reference feature maps, including ACC and VEL), followed by the application of the $Softmax$ activation function for normalisation to derive the attention weights. Consequently, the components related to the MA feature map are removed to obtain a noise-free PPG feature map through \eqref{e4}.

Meanwhile, the `multi-head' in the proposed attention mechanism signifies that this attention computation is performed multiple times in parallel, with each 'head' learning different attention patterns, thereby capturing different types of relationships between the encoded PPG and motion reference (ACC and VEL) feature maps. In this work, we considered that the number of heads is eight  for the proposed attention mechanism (i.e., $h=8$.)

\begin{algorithm}[H]
\caption{Multi-sensor fusion cross-attention mechanism pseudocode}
\begin{spacing}{1.3}
\begin{algorithmic}[1]
\Statex \(\#\) encoder network $g_{enc}$, decoder network $g_{dec}$
\Statex \(\#\) query $Q$, key $K$ and value $V$
\State $g_{enc}.\text{params}, g_{dec}.\text{params}$  \Comment{initialise encoder and decoder}

\For{\textbf{each} minibatch $x_{PPG}, x_{ACCVEL}$ \textbf{in} \textit{loader}}
    \State $x_{PPG}, x_{ACCVEL} \gets \mathsf{preprocess}(x_{PPG}),\; \mathsf{preprocess}(x_{ACCVEL})$

    \State $X_{PPG}, X_{ACCVEL} \gets g_{enc}(x_{PPG}),\, g_{enc}(x_{ACCVEL})$ \Comment{encoder outputs}
    \State $V, K \gets conv_{1\times1}(X_{PPG})$ \Comment{$V, K$ outputs}

    \State $Q \gets conv_{1\times1}(X_{ACCVEL})$ \Comment{$Q$ outputs}

    \State $W_{MA} \gets softmax((QK^T)/\sqrt{d_k})$  \Comment{MA attention weights}
    \State $S \gets X_{PPG}-W_{MA}V$   \Comment{MA-removal signal features}
    \State $s_{PPG} \gets g_{dec}(S)$ \Comment{MA-removal PPG signal}
     
\EndFor
\end{algorithmic}
\end{spacing}
\label{algo1}
\end{algorithm}

\subsection{Discriminator for Selective Inputs }
\begin{figure}[!hbt]
    \centering
    \includegraphics[width=0.8\textwidth]{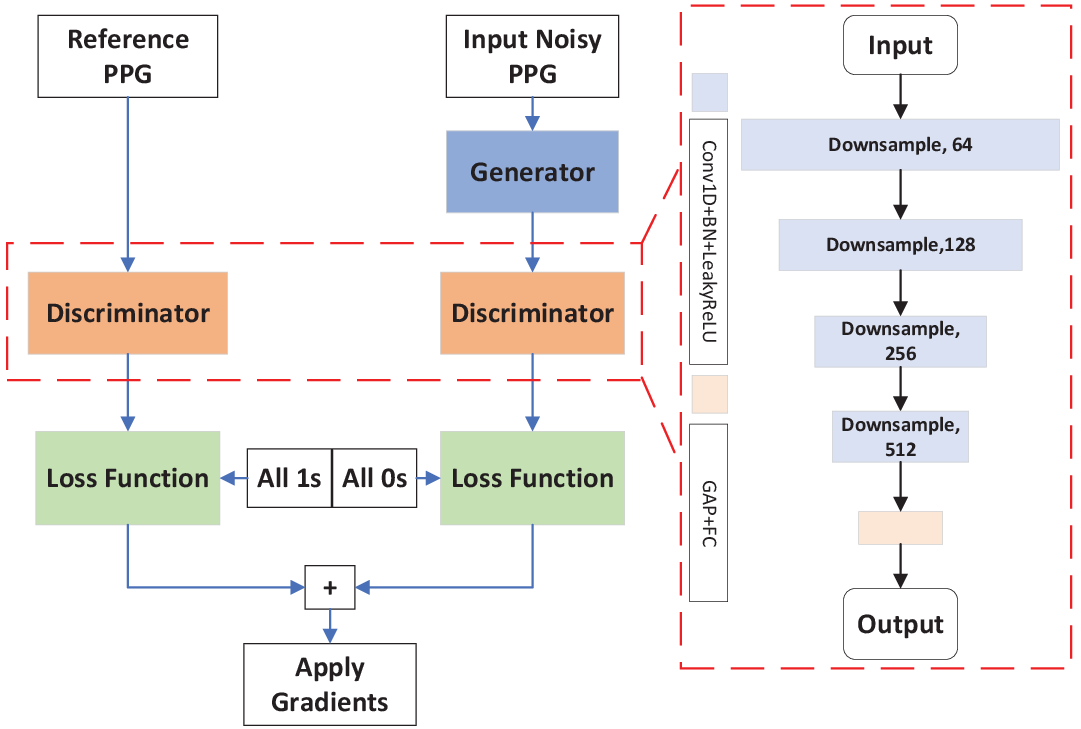}
    \caption{The discriminator structure.}
    \label{Fig4}
\end{figure}

The Discriminator is a network structure with four 1D-convolutional layers followed by a global average pooling (GAP) layer and a fully connected (FC) as the final layer (illustrated in Fig.~\ref{Fig4}). The discriminator adopts the leaky rectified linear (LeakyReLU) activation function and incorporates batch normalisation to accelerate convergence. The discriminator evaluates probabilities, specifically assessing the probability of the output of the generator, $D(G(p(t), a_{x,y,z}(t)))$, and the probability associated with the reference signal, $D(s_{ref}(t))$.

\subsection{Loss function}
Noise removal in the PPG signal can be expressed by the equation $p(t)=s(t)+n(t), t \geq 0$ where $p(t)$ represents the raw PPG signal that includes MAs, $s(t)$ denotes the target PPG signal, and $n(t)$ is the noise that needs to be eliminated. In the discrete case
\begin{equation}
\label{e5}
s[i]=p[i]-n[i], i=1,2,3...
\end{equation}
where $i$ is the sampling index, and the $\pmb{p}=[p[i]]$, $\pmb{n}=[n[i]]$ and $\pmb{s}=[s[i]]$ are all 1D vectors. By capitalising on its strong ability to distinguish between noisy PPG signals, $\pmb{p}$, and reference PPG signals, $\pmb{s}_{ref}$, the method effectively learns the distribution features of both. This capability allows it to generate the de-noised PPG signal, $G(\pmb{p},\pmb{a}_{x,y,z})$.  
The waveform error loss, Mean Squared Error ($L_{MSE}$), is defined as calculating the point-wise error between $\pmb{s}_{ref}$ and $G(\pmb{p},\pmb{a}_{x,y,z})$ in the time domain, is given by
\begin{equation}
\label{e6}
L_{MSE}=\frac{1}{n} \sum_{i=1}^{n} \left \|\pmb{s}_{ref} - G(\pmb{p, a}_{x,y,z})\right\|_{2}
\end{equation}
where $\left \| \cdot \right\|_{2}$ represents the $L2$ norm. 

In summary, the overall loss function for the generator can be defined as:
\begin{equation}
\label{e8}
L_{G}= \log(1-D(G(\pmb{p, a_{x,y,z}}))+\lambda L_{MSE} 
\end{equation}
where the first item denotes the binary cross-entropy loss, and $\lambda$ is the balancing weight. In our experiments, the $\lambda$ is set to 1000 empirically based on the individual loss scales.

The discriminator binary cross entropy loss function is~\cite{b20ronneberger2015u} 
\begin{equation}
\label{e9}
\begin{split}
L_{D} =\log(D(\pmb{s}_{ref}))+\log(1-D(G(\pmb{p, a_{x,y,z}})))
\end{split}
\end{equation}
with which the discriminator differentiates between the generated and reference PPG signals. Through the adversarial learning process between the generator and the discriminator, the generator consistently acquires both global and local features. This learning ensures that the PPG signal produced closely mirrors the characteristics of the reference signal.

\section{Selective Measurement Protocols}
Four datasets provided by Loughborough University (in-house LU dataset), the IEEE Signal Processing Cup 2015 (IEEE-SPC dataset) and the publicly available PPG dataset (PPGDalia dataset) with their ethical approvals are experimented with the proposed AM-GAN.

The in-house LU dataset was collected with a multi-wavelength illumination optoelectronic patch sensor (mOEPS) at a sampling rate of 256Hz. It includes data from 12 subjects, aged between 23 and 37 years, recorded while they exercised on a treadmill \cite{b21zheng2023rapid}. The subjects engaged in exercise at four varying intensities, including 3$km/h$, 6$km/h$, 9$km/h$, and 12$km/h$, with each stage lasting four minutes and a 60-second recovery period between them. Additionally, the subjects engaged in cycling at a constant rate of 60 Revolutions Per Minute (RPM) across four different intensity levels, with power outputs set at 60 W, 100 W, 120 W, and 150 W. Each stage lasted four minutes with a 60-second recovery break in between. A ground-truth HR was captured using a Polar Bluetooth$^\circledR$ Smart chest strap (Polar Electro, Kempele, Finland), and a ground-truth RR was measured using a Vyntus$^\text{TM}$ CPX Metabolic cart (JAEGER$^\text{TM}$ Vyntus$^\text{TM}$ CPX, Carefusion, Germany).

The in-house multi-wavelength C2 datasets were collected using a protocol implemented in a controlled environmental chamber, where measurements (at rest and in motion) were taken at specific simulated altitudes at three different Oxygen levels of 21\%, 18\% and 16\%. During cycling exercise, the subjects performed the exercise at three different intensities; each stage lasted four minutes with a recovery of 60 seconds in between. The data was simultaneously gathered from the multi-wavelength C2 devices at a sampling frequency of 1024 $Hz$ worn on the wrist and attached on the chest, from four subjects aged  25 - 35 years and with type I, type II, type IV, and type VI skin tones (Fitzpatrick scale). In the cycling exercise, continuous SpO$_2$ reference data were obtained using a Finapres$^\circledR$ NOVA patient monitor.

The IEEE-SPC dataset features a two-channel PPG signal using green (515 nm) LEDs, a triaxial ACC signal, and an ECG reference, collected from 12 male subjects aged between 18 and 35 years. These datasets were sampled at a frequency of 125 Hz \cite{b6zhang2014troika}. The IEEE-SPC includes the training and testing datasets. In the training datasets, 12 subjects ran at varying speeds. Furthermore, 10 subjects (aged 19-58 years) performed intensive arm movements, i.e., boxing, in the testing datasets.

The PPGDalia includes synchronised recordings of ECG, wrist-worn PPG, and acceleration data from 15 subjects. Each subject contributed approximately two hours of recorded data. A consistent time shift between PPG and triaxial ACC signals was identified and manually fixed \cite{b29reiss2019deep}.

\section{Experimental Results}
The experimental studies constituting the three listed datasets were carried out to obtain physiological measurements allowing the extraction of two types of physiological parameters, i.e., HR and RR. These were used to evaluate the performance of the proposed AM-GAN with three physical activity datasets comprising: 1) network ablation experiments to assess the effectiveness of the proposed method; 2) performance evaluation of the proposed AM-GAN against some other recently reported methods; 3) effectiveness examination of the proposed AM-GAN in handling cross-dataset physiological parameter estimation. 

\subsection{Experimental Setup}
An eight-second sliding window technique is applied to analyse both the input and reference PPG signals, advancing one second at each step. All input and reference PPG signals are downsampled to 32 $Hz$. During the experiments, the proposed AM-GAN undergoes training for 100 epochs, with both the generator and discriminator networks being optimised concurrently using the Adam optimiser at a learning rate of 0.0002. The input PPG data is pre-processed with an eighth-order Butterworth band-pass filter with band edges set at 0.2 $Hz$ and 6.5 $Hz$. We apply one-sided label smoothing to the discriminator’s real targets: during training, the ground-truth label for real images is set to 0.9 instead of 1.0. This small perturbation prevents the discriminator from becoming over-confident and keeps its gradients in a meaningful range. In all experiments, HR and RR are computed using the extraction algorithms defined in \cite{b21zheng2023rapid}.

\subsection{Experimental Results} 
\begin{figure}[!hbt]
 \vspace{-0.5cm}
    \centering
    \setlength{\belowcaptionskip}{-0.3cm} 
    \begin{minipage}[b]{1\textwidth}
        \subfloat[]{
            \includegraphics[width=0.48\textwidth]{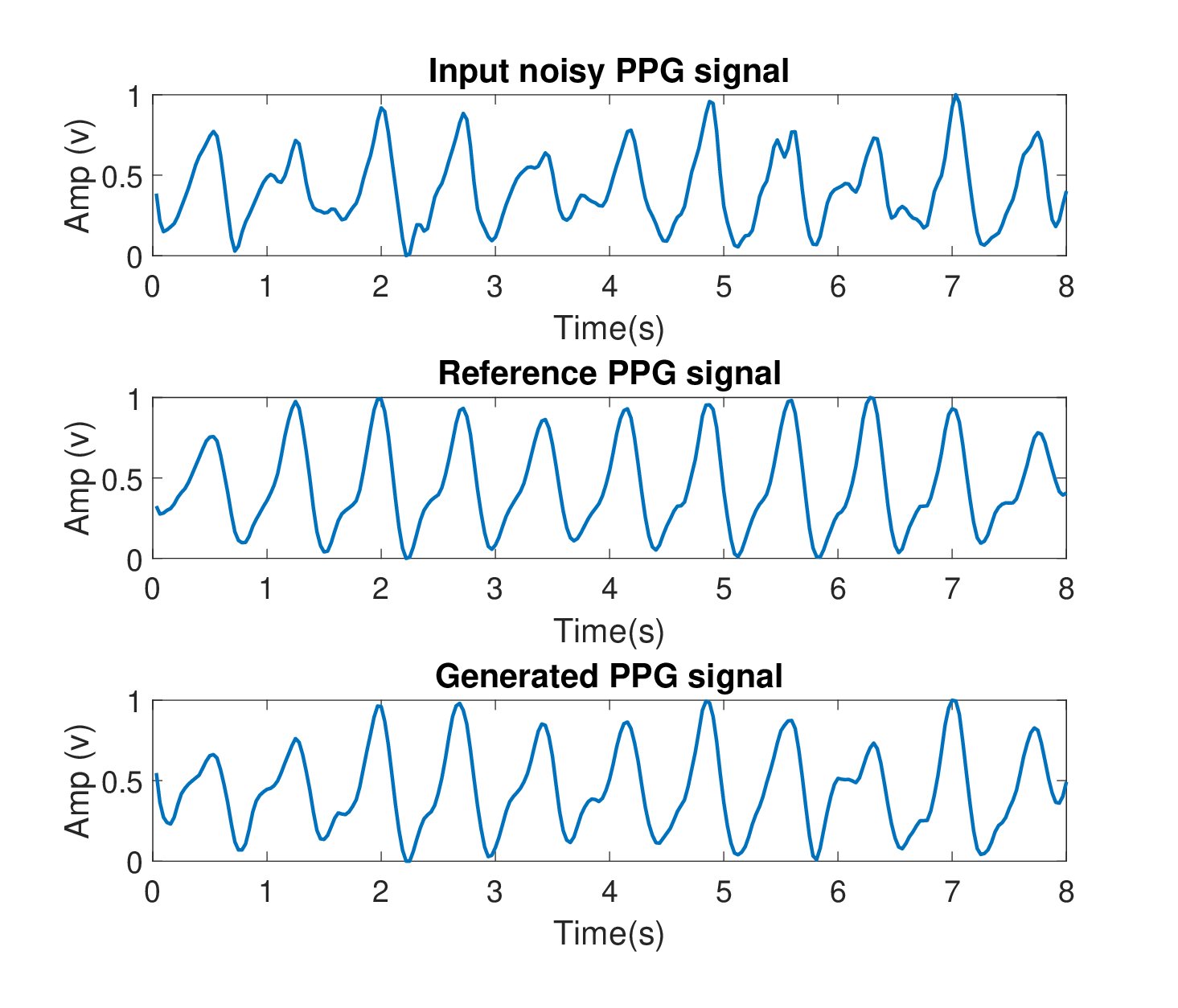}
            \label{label_for_cross_ref_1}
        }\hspace{0mm}
        \subfloat[]{
    	\includegraphics[width=0.48\textwidth]{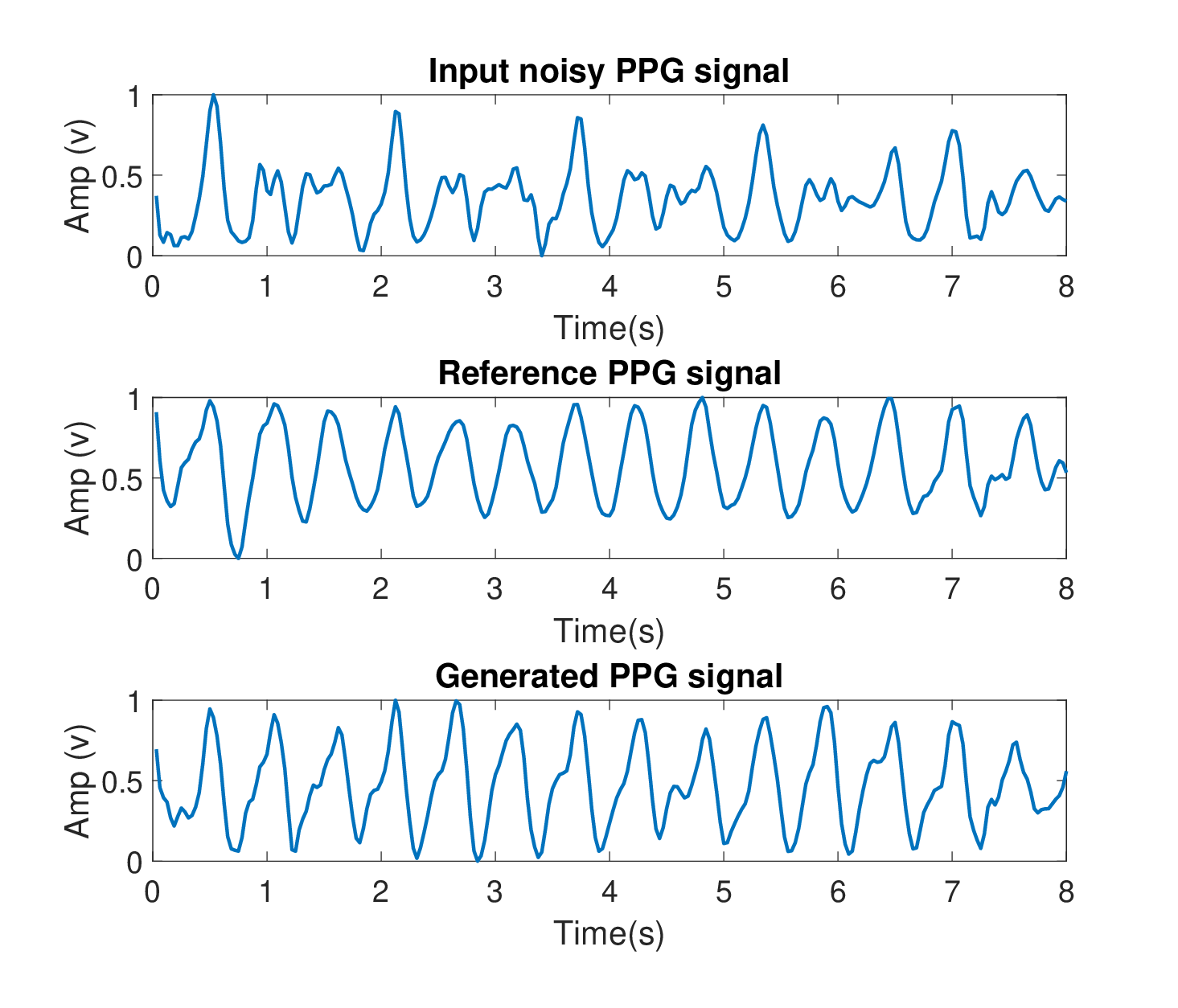}
            \label{label_for_cross_ref_2}
        }\hspace{0mm}
        \quad    
        \subfloat[]{
        	\includegraphics[width=0.48\textwidth]{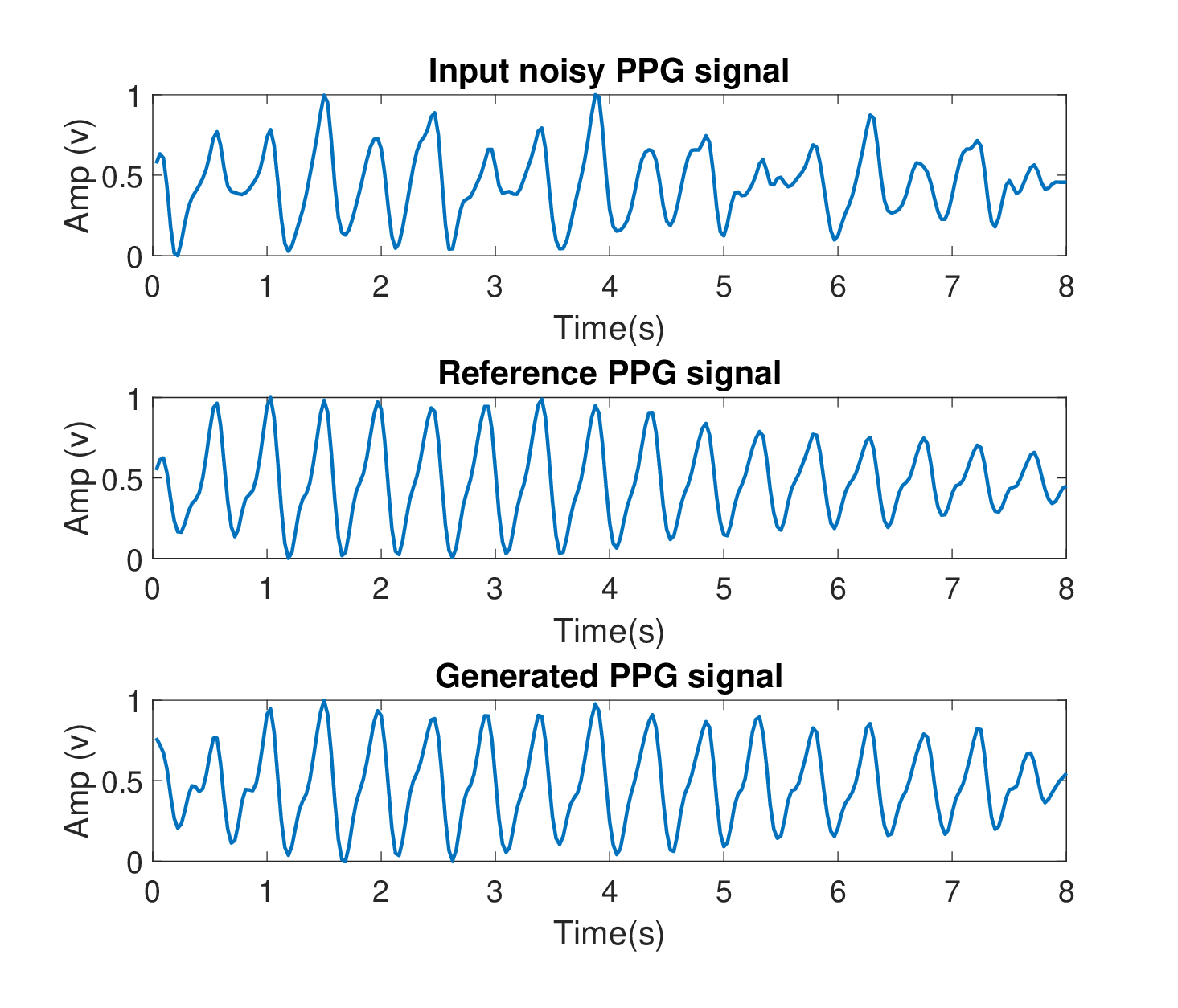}
            \label{label_for_cross_ref_3}
        }\hspace{0mm}
        \subfloat[]{
    	\includegraphics[width=0.48\textwidth]{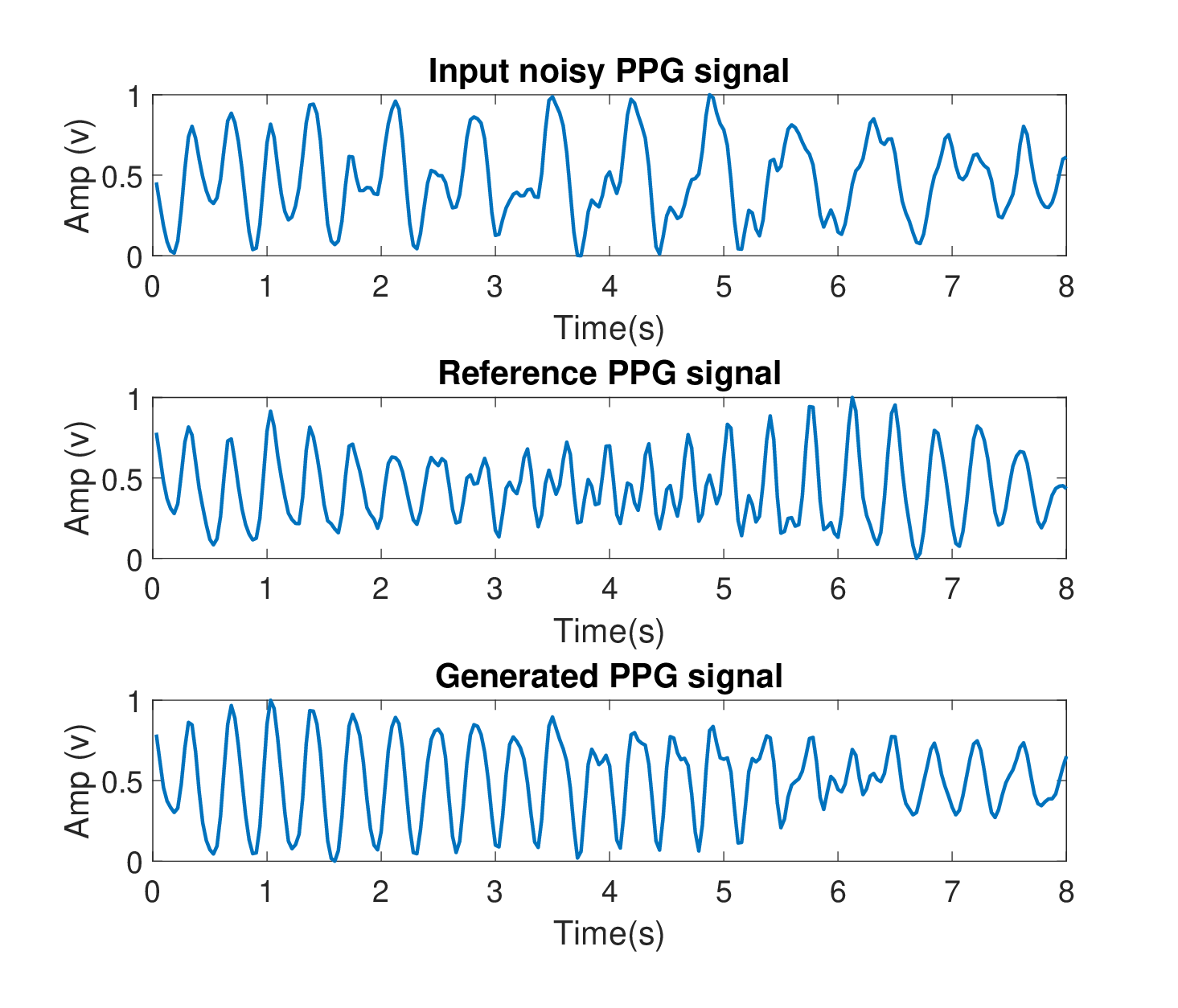}
            \label{label_for_cross_ref_4}
        }
    \end{minipage}
    \caption{A representative example for the waveform comparison of the original noisy signal, the reference signal and the generated signal at different physical activity intensities; i.e., (a) at rest, (b) low-intensity activity (6$km/h$), (c) medium-intensity activity (9$km/h$), and (d) high-intensity activity (12$km/h$).}
    \label{Fig14}
\end{figure}

\textbf{Waveform quality on in-house LU dataset:} Fig. \ref{Fig14} shows the selected PPG waveform in the time domain from the different physical activity intensities (subject-M10, treadmill), i.e., rest, low-intensity (6$km/h$), medium-intensity (9$km/h$), and high-intensity (12$km/h$). The generated PPG signals illustrate that the AM-GAN is capable of yielding results that closely approximate the reference signals across various levels of exercise intensity, thereby achieving satisfactory outcomes.

\begin{figure}[hbt]
    \setlength{\belowcaptionskip}{-0.3cm} 
    \begin{minipage}[b]{1\textwidth}
    \centering
    \subfloat[]{
        \includegraphics[width=0.48\textwidth]{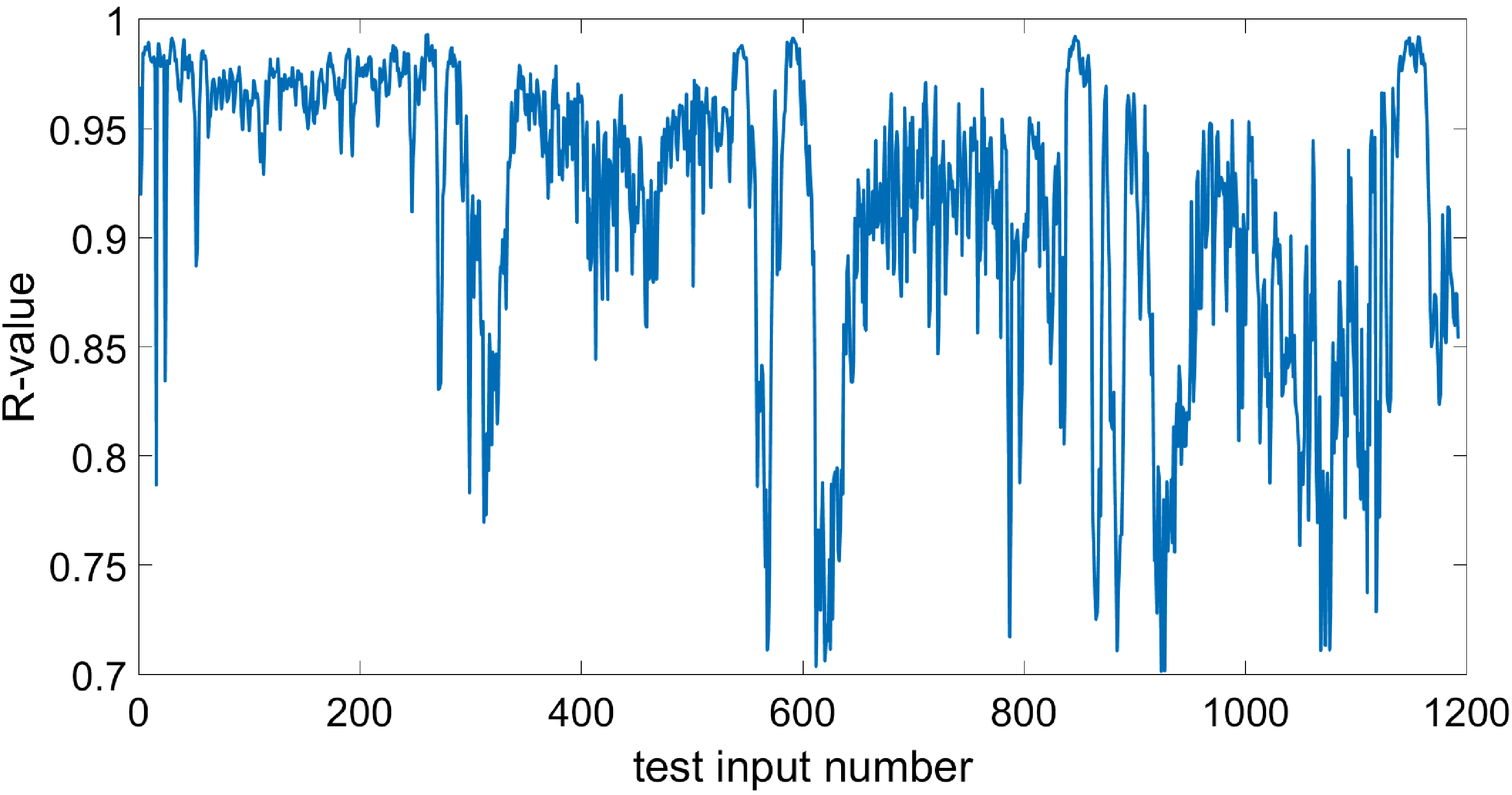}
    }\hspace{-3mm}
    \subfloat[]{
        \includegraphics[width=0.48\textwidth]{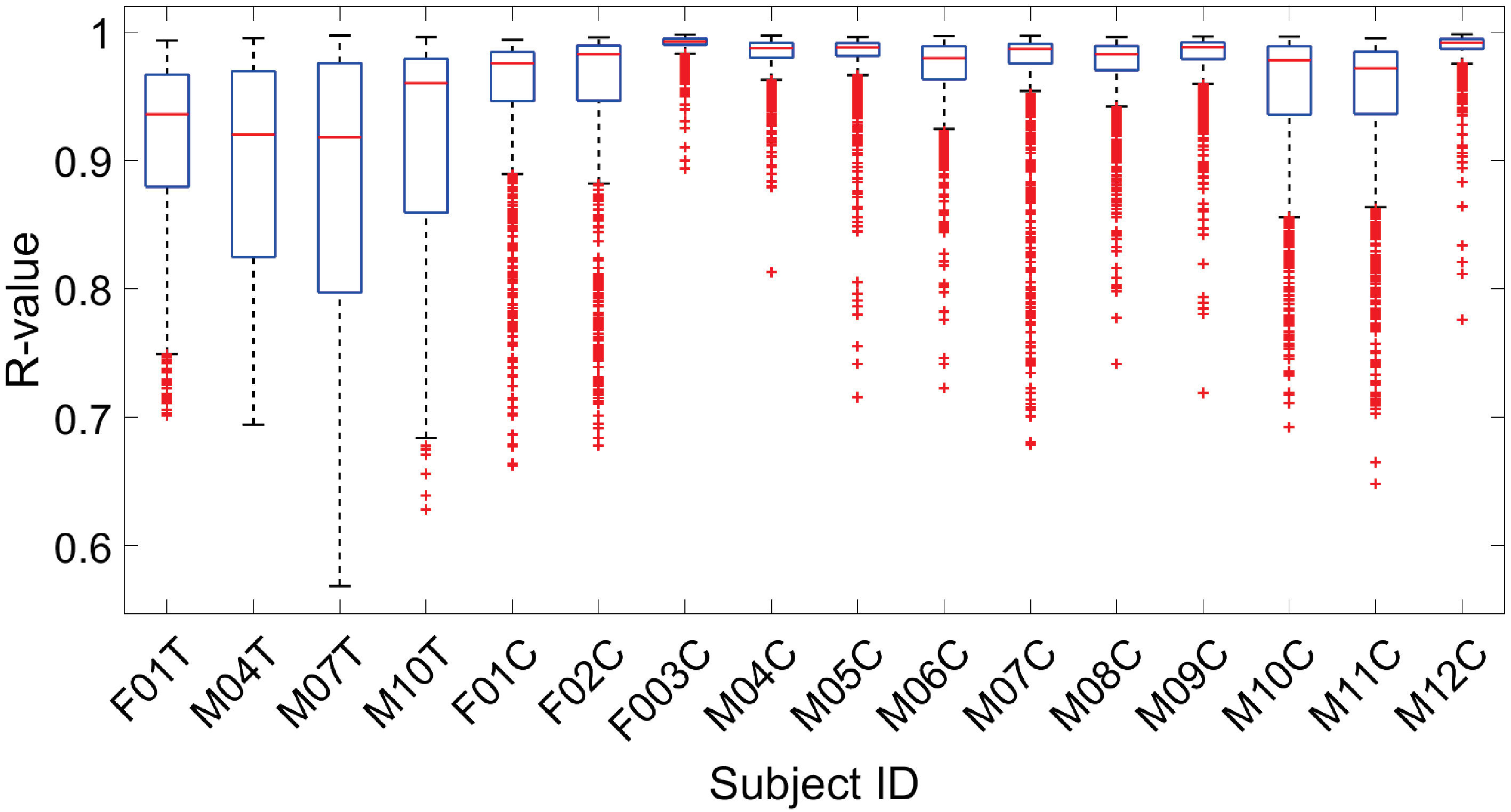}
    }
    \caption{Waveform quality validation on LU subjects (F-female, M-male, T-treadmill, C-cycling). (a) Waveform quality ($R$) for F01-treadmill. (b)Mean waveform quality ($\overline{\langle R \rangle}$) across all LU testing subjects.}
    \label{Fig13}
    \end{minipage}
\end{figure}

\begin{equation}
\label{e10}
R= \frac{\int_{0}^{T} G(p(t), a_{x,y,z}(t))s_{ref}(t)dt}{\sqrt{\int_{0}^{T} G(p(t), a_{x,y,z}(t))^2dt} \sqrt{\int_{0}^{T} s_{ref}(t)^2dt}}
\end{equation}
To validate the error of the PPG signal generated by the AM-GAN method with respect to the reference PPG signal, the Pearson correlation ($R$) was applied for quantifying waveform, as shown in \eqref{e10}, where $G(p(t),a_{x,y,z}(t))$ is the zero-mean generated PPG signal and $s_{ref}(t)=MR(p(t),a_{x,y,z}(t))$ is the zero-mean reference signal, for each vector $p(t)$ in the test set, $\mathcal{P}_{Te}$. An eight-second sliding time window was employed for the calculation of the waveform quality. The waveform quality index ($R$), for the treadmill exercise test cases of subject F01, is shown in Fig. \ref{Fig13}(a). The mean Pearson correlation ($\overline{\langle R \rangle}$) for all LU testing subjects is illustrated in Fig. \ref{Fig13}(b). It is clear that the AM-GAN outputs are highly correlated with the reference signals with a high level of consistency across the entire set of test vectors and for all testing subjects, and the $Mean$ value of all testing subjects' waveform quality ($\overline{\langle R \rangle}$) is 0.9522.

\textbf{HR and RR calculation on in-house LU dataset:} To evaluate the effectiveness of the method, heart rate (HR) and respiratory rate (RR) are independently computed for both the generated signals $G(p(t),a_{x,y,z}(t))$ and the reference PPG signals $s_{ref}(t)$. These results are then compared with readings from commercial devices using the in-house dataset. Following the in-house protocol, the first eight subjects are used for training and validation with a ratio of 0.7 and 0.3, adopting the green-channel illumination during the treadmill exercise and the remaining four subjects are used for testing. Also, data from these 12 subjects during cycling exercise at different intensities is applied in the procedure.

\begin{figure}[hbt]
    \setlength{\belowcaptionskip}{-0.3cm} 
    \begin{minipage}[b]{1\textwidth}
    \centering
    \subfloat[]{
        \includegraphics[width=0.48\textwidth]{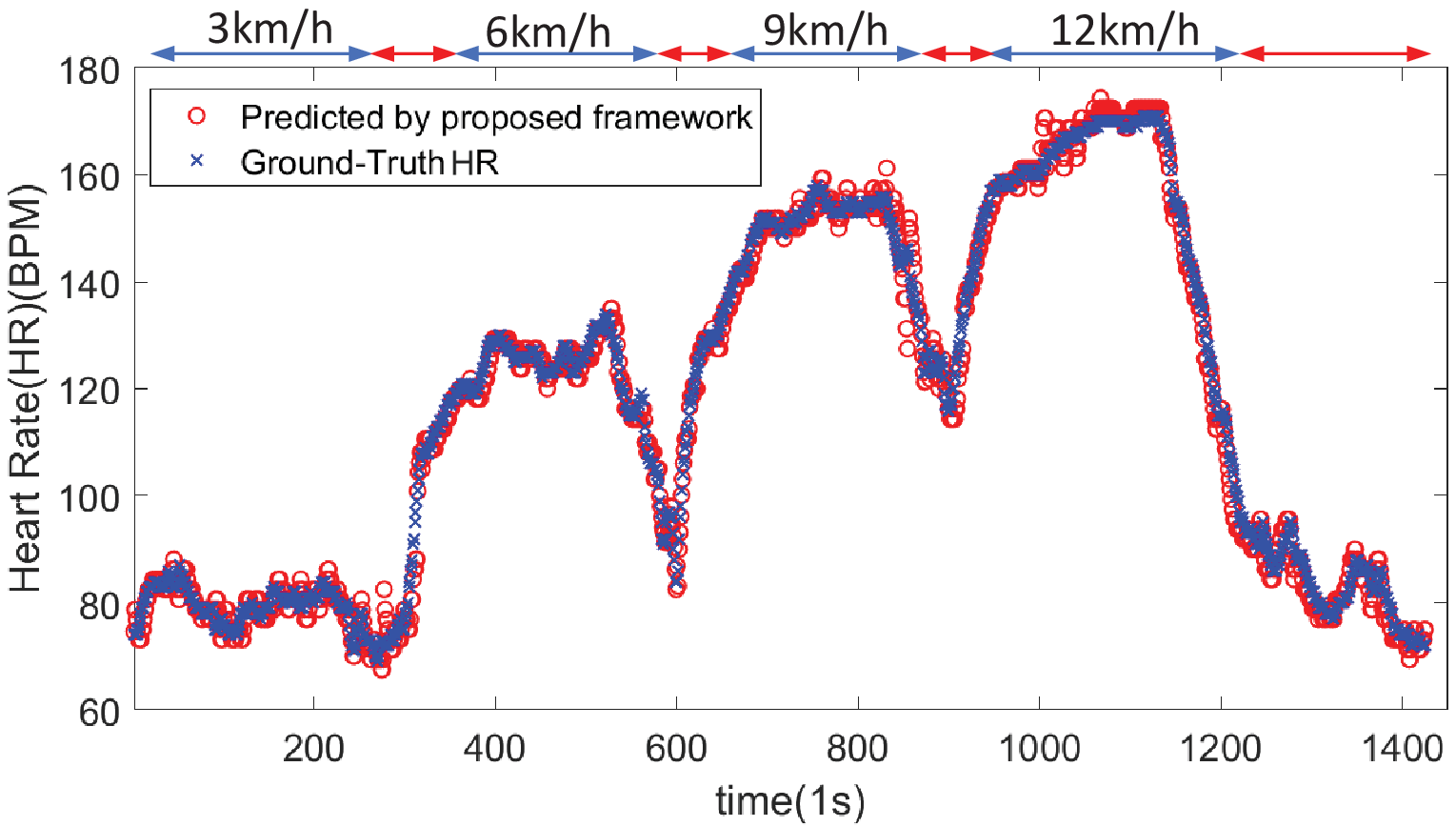}
    }\hspace{-3mm}
    \subfloat[]{
        \includegraphics[width=0.48\textwidth]{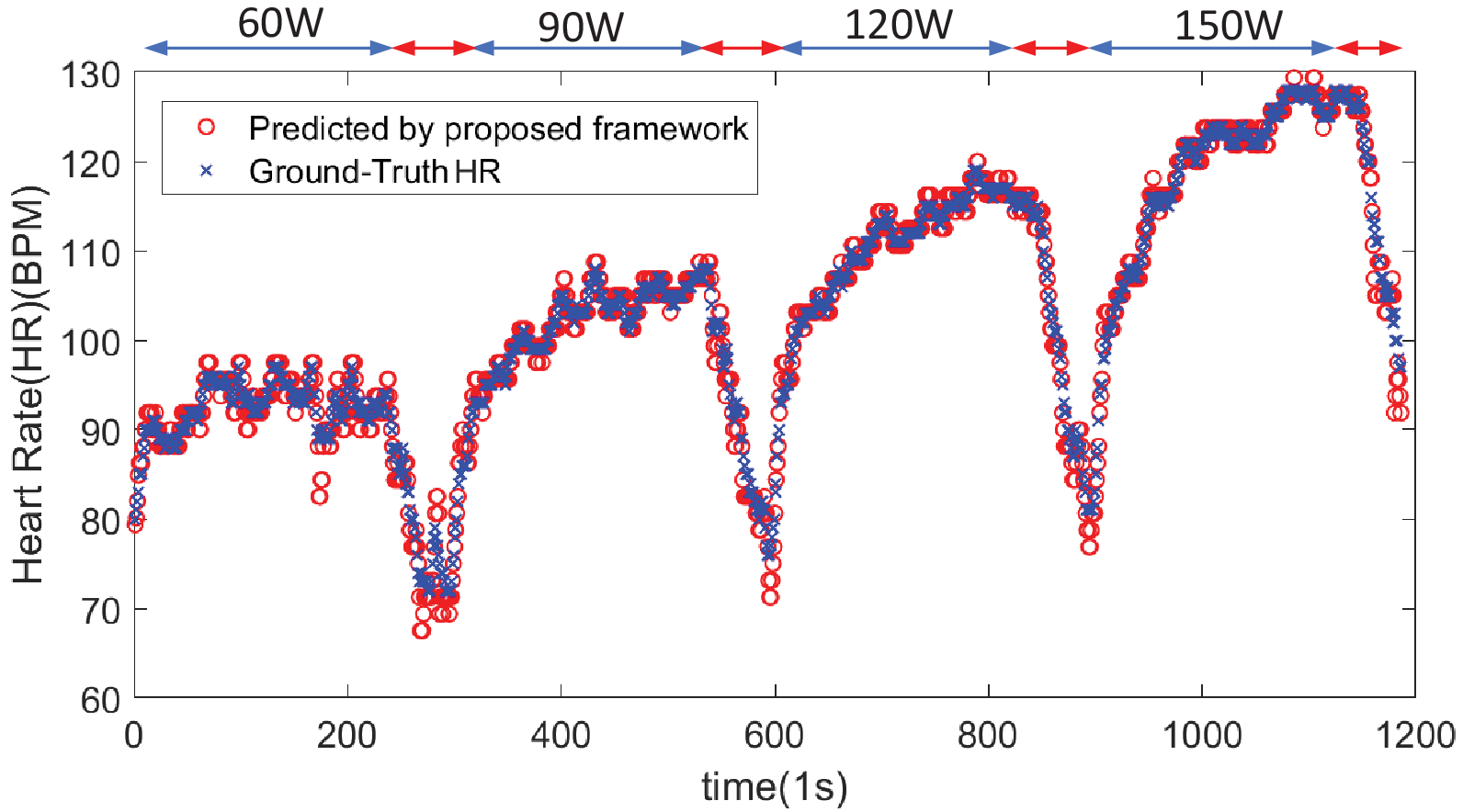}
    }
    \caption{HR calculation outcomes for two randomly selected datasets (in BPM: beats/min) are displayed. (a) Shows the results for the M10 dataset during treadmill exercise, while (b) shows results for the M08 dataset during cycling exercise. The top of the figure indicates the various exercise intensities, with the red arrow pointing to the rest status.}
    \label{Fig5}
    \end{minipage}
\end{figure}

The outcomes of continuous HR monitoring during treadmill and cycling exercises are shown in Fig.~\ref{Fig5}(a) and \ref{Fig5}(b), where they are compared to the reference values from a commercial device. In Fig.~\ref{Fig5}, \textit{Polar} refers to the reference heart rate recorded with a Polar Bluetooth Smart chest strap. In both figures, the MA-removal reference uses our previously published algorithm~\cite{b21zheng2023rapid}. The accuracy in capturing HR is consistent and closely matches the provided reference.

\begin{figure}[hbt]
    \setlength{\belowcaptionskip}{-0.3cm} 
    \begin{minipage}[b]{1\textwidth}
    \centering
    \subfloat[]{
        \includegraphics[width=0.48\textwidth]{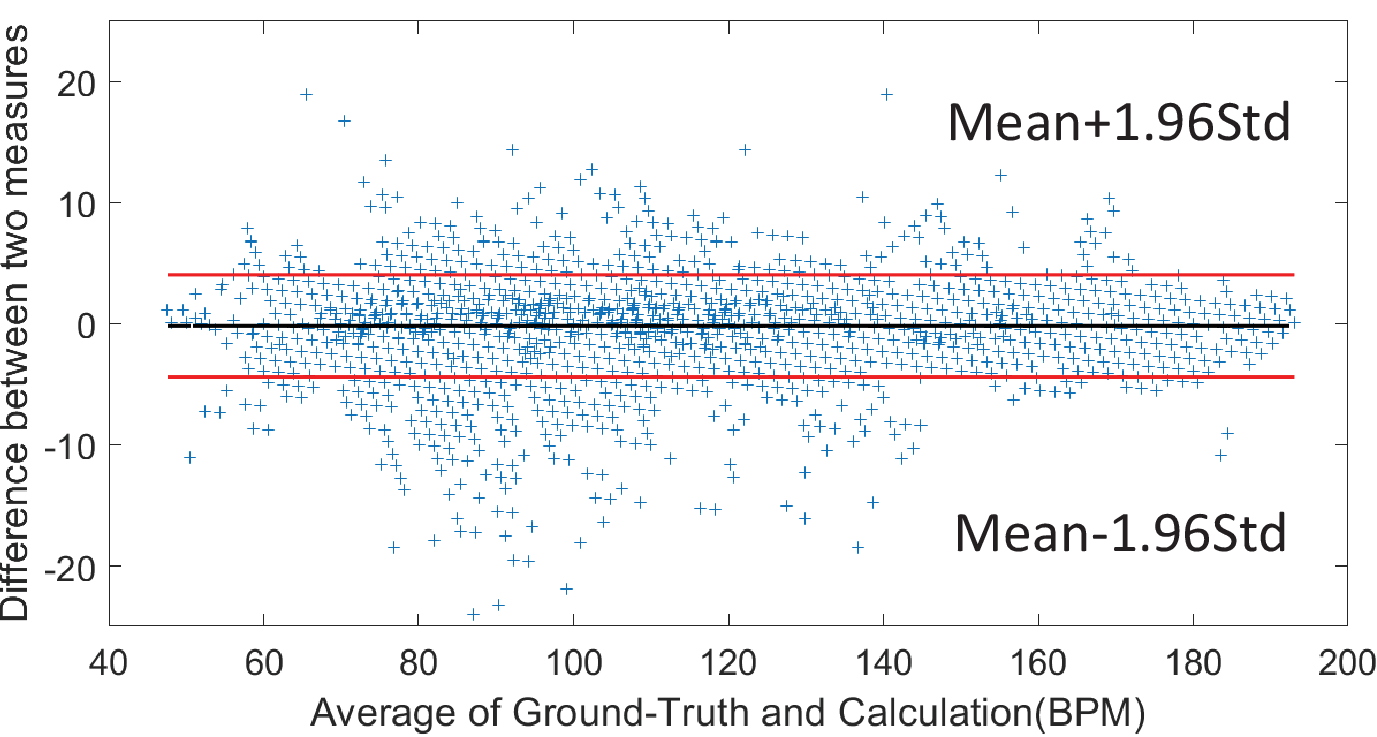}
    }\hspace{-3mm}
    \subfloat[]{
        \includegraphics[width=0.48\textwidth]{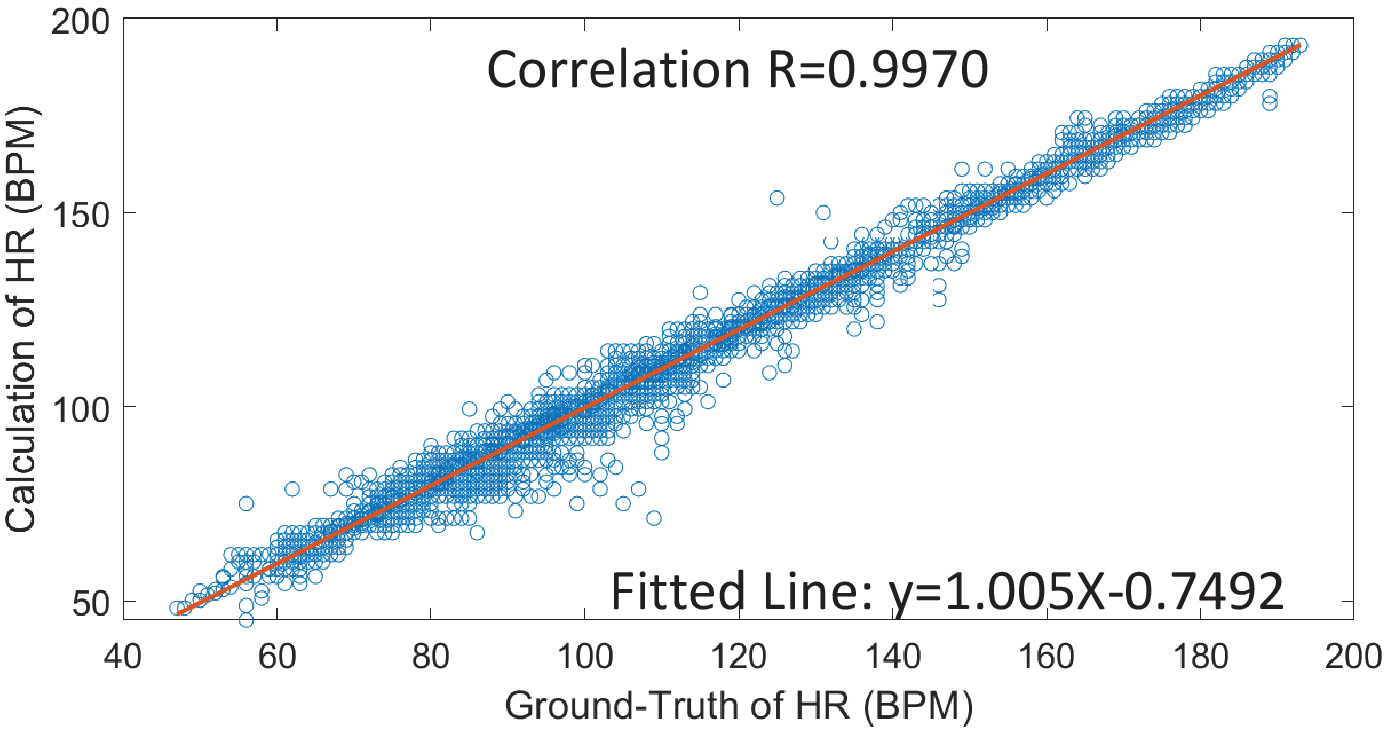}
    }
    \caption{HR results calculated from the 16 testing subject datasets (in BPM: beats/min) in the cycling and treadmill exercises. (a) Bland-Altman plot. (b) Scatter plot of Pearson correlation.}
    \label{Fig6}
    \end{minipage}
\end{figure}

The Bland-Altman plot comparing the ground truth to the computed outcome is depicted in Fig. \ref{Fig6}(a). In this case, the limits of agreement (LOA) ranged from $[-4.04, 3.80]$ beats/min, with 95\% of the differences falling within this interval. Fig. \ref{Fig6}(b) shows the scatter plot comparing the ground-truth HR values to those calculated for 16 test subjects (12 undergoing cycling and 4 on treadmill exercises). Additionally, the line of best fit, represented as 
$y=1.005x-0.7492$, is displayed, where $x$ is the ground-truth HR and  $y$ is the calculated HR. The Pearson correlation coefficient $R$ achieved using the proposed AM-GAN is 0.9970.

\begin{figure}[!hbt]
    \setlength{\belowcaptionskip}{-0.3cm} 
    \begin{minipage}[b]{1\textwidth}
    \centering
    \subfloat[]{
        \includegraphics[width=0.48\textwidth]{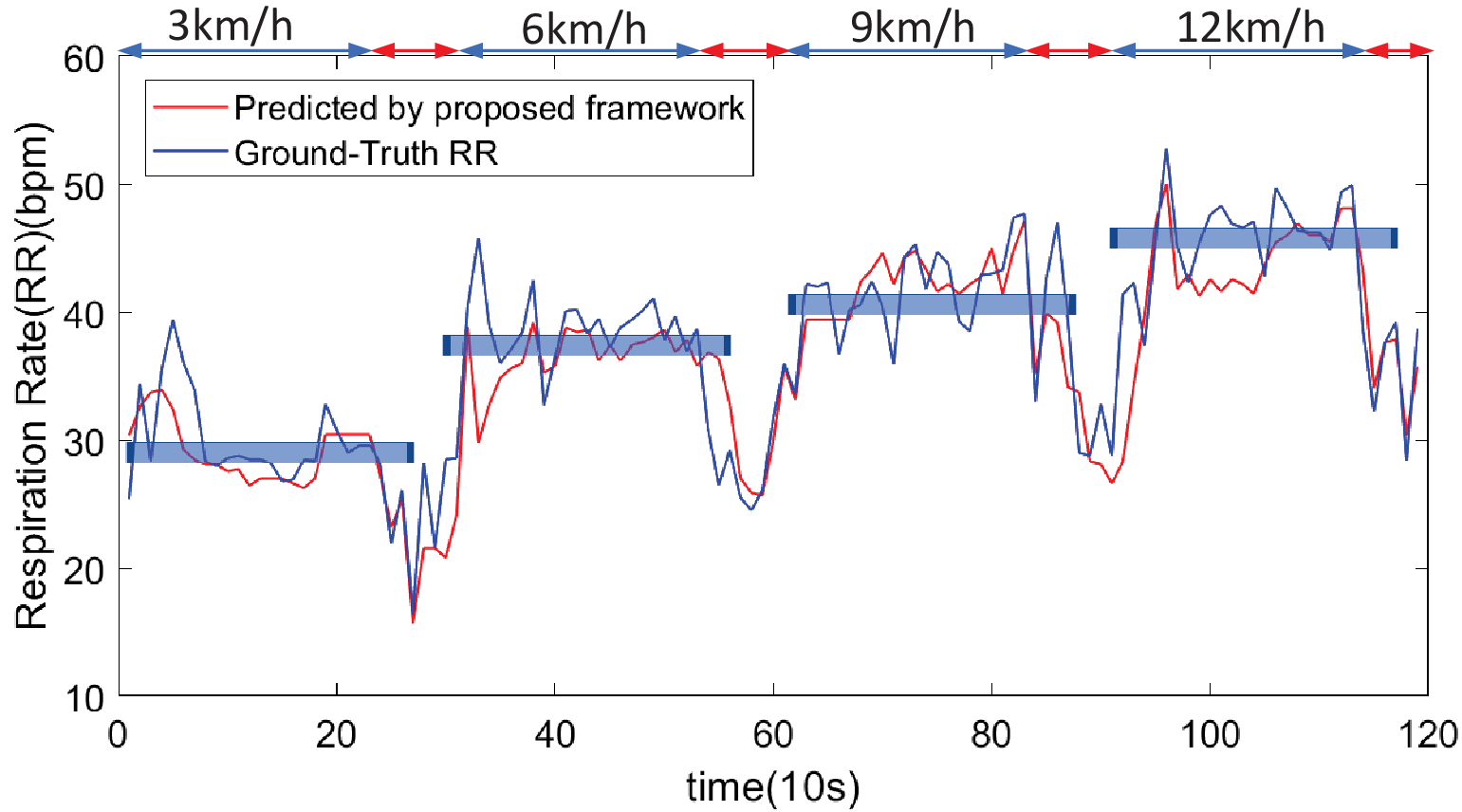}
    }\hspace{-3mm}
    \subfloat[]{
        \includegraphics[width=0.48\textwidth]{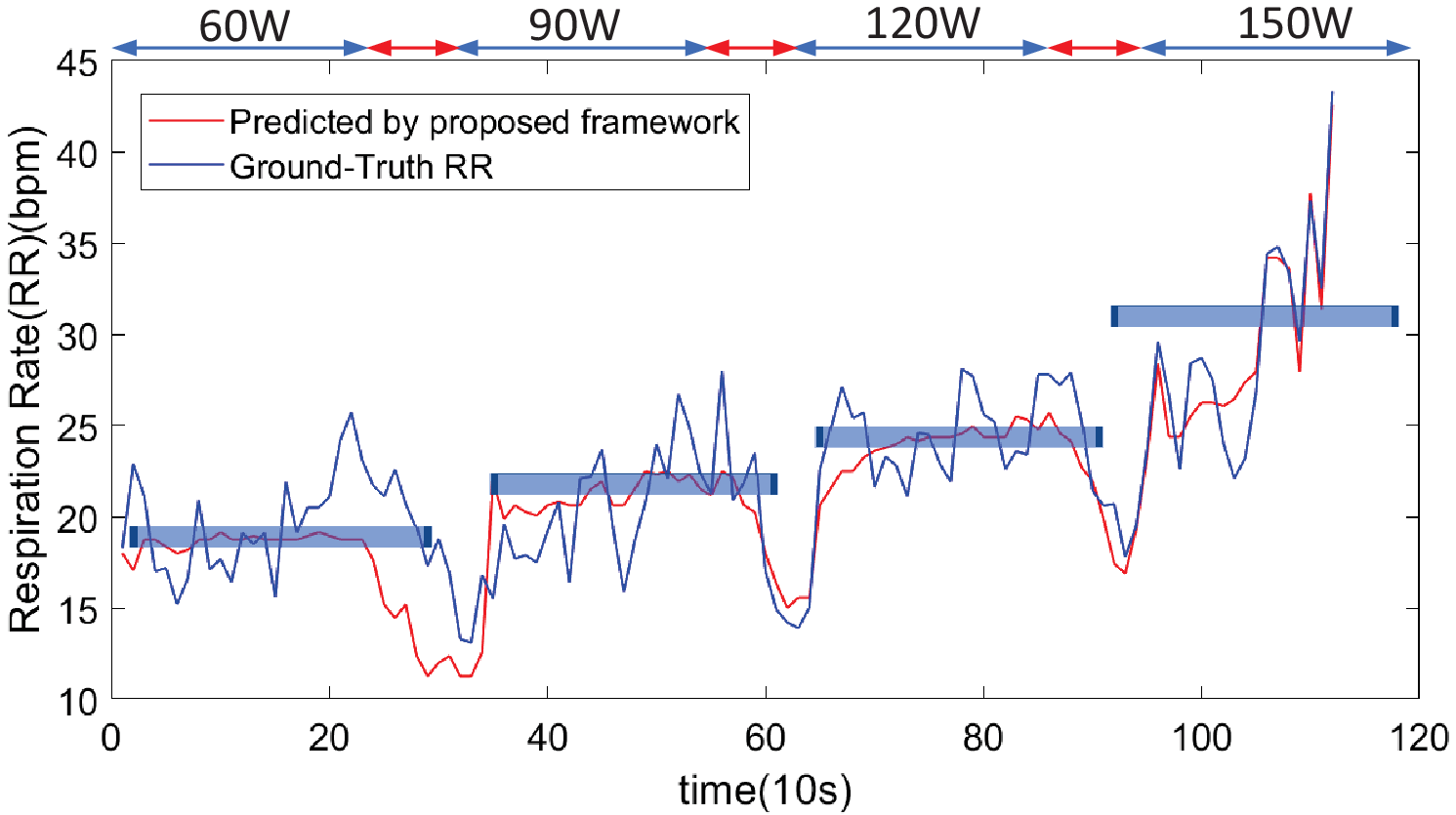}
    }
    \caption{RR calculation results on two randomly selected datasets (in breaths/minute, bpm). (a) Results from the F01 dataset during treadmill exercise. (b) Results from the M04 dataset during cycling exercise. The top of the figure indicates the various exercise intensities, with the red arrow pointing to rest periods.}
    \label{Fig7}
    \end{minipage}
\end{figure}

\begin{figure}[hbt]
    \setlength{\belowcaptionskip}{-0.3cm} 
    \begin{minipage}[b]{1\textwidth}
    \centering
    \subfloat[]{
        \includegraphics[width=0.48\textwidth]{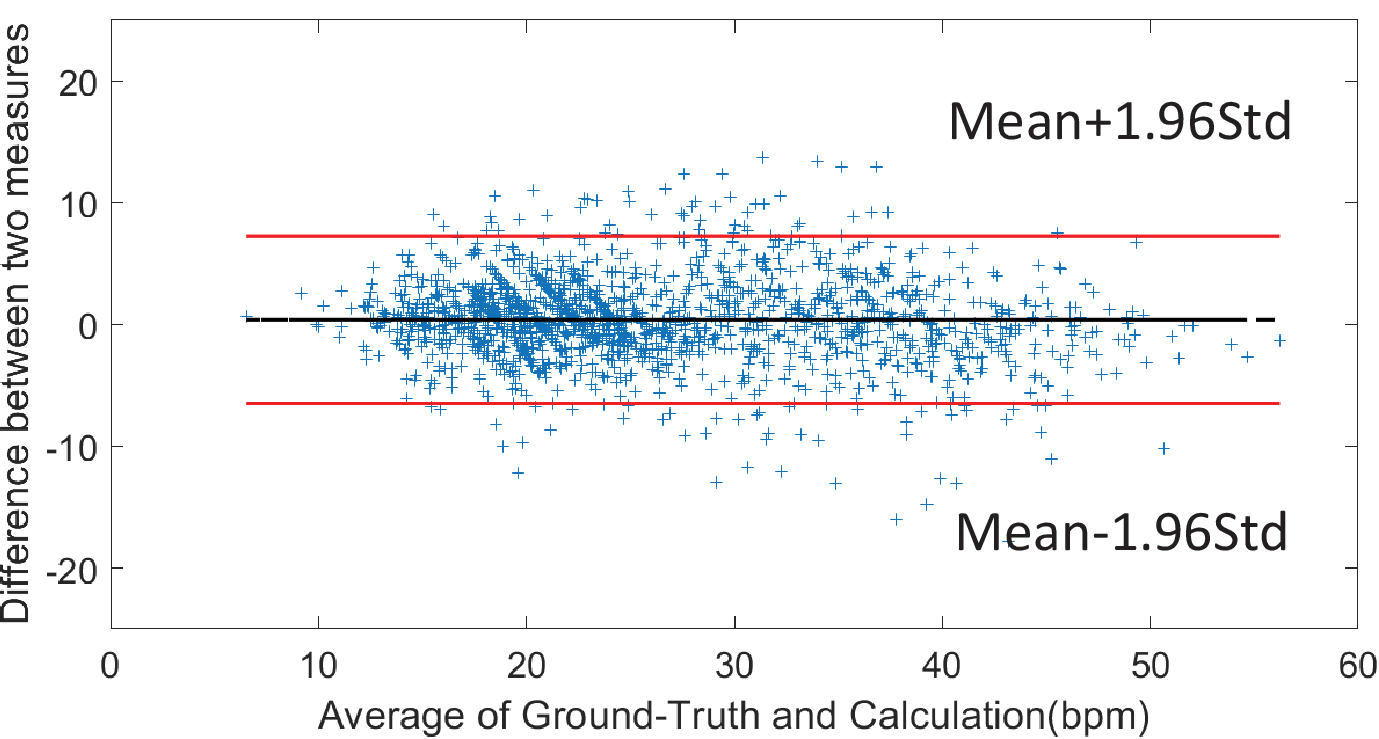}
    }\hspace{-3mm}
    \subfloat[]{
        \includegraphics[width=0.48\textwidth]{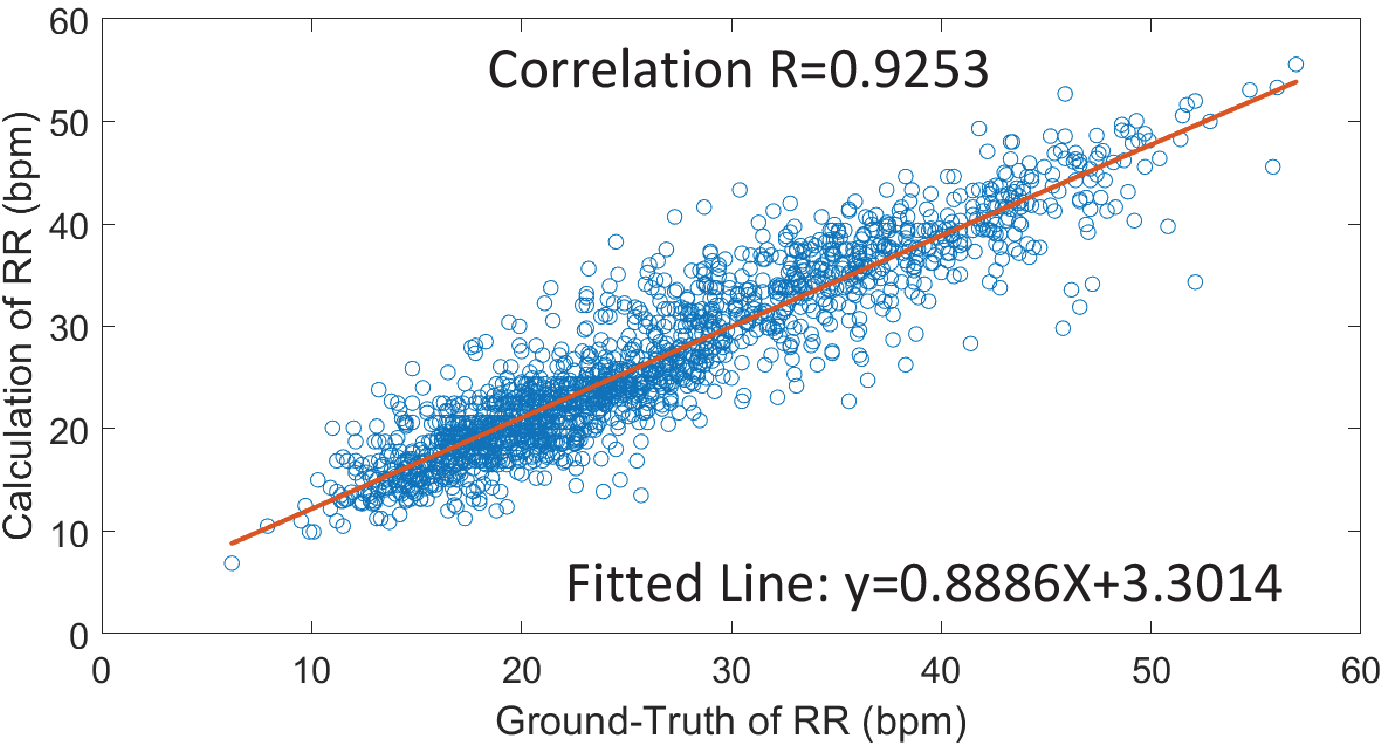}
    }
    \caption{RR results, measured in breaths/minute (bpm), were derived from the datasets of 16 test subjects during cycling and treadmill exercises. Displayed are (a) a Bland-Altman plot and (b) a scatter plot illustrating the Pearson correlation.}
    \label{Fig8}
    \end{minipage}
\end{figure}

The RR performance in the presence of MAs is demonstrated in Figs. \ref{Fig7} and \ref{Fig8}. Fig. \ref{Fig7} presents the RR derived from two randomly selected datasets. The blue line represents the computed RR, while the red line indicates the true RR as measured by the Vyntus$^\text{TM}$ CPX Metabolic cart. As illustrated in Fig.\ref{Fig7}, the computed RR closely aligns with the structure of the ground-truth RR. Furthermore, due to significant fluctuations in the continuously measured RR shown in Fig.\ref{Fig7}, a blue-shaded area is included to depict the average RR across various exercise stages. This provides a more intuitive representation of changes in the RR.

The Bland-Altman plot for all subjects is presented in Fig. \ref{Fig8}(a). The LOA between the ground-truth and the calculated RR data is [-6.48, 7.26] breaths/min, and 95\% of all differences were inside this range. Furthermore, a fit line $y=0.8886x+3.3014$ has been constructed to represent the relationship between the ground-truth and calculated RR, demonstrating a Pearson correlation coefficient ($R=0.9253$), as illustrated in Fig. \ref{Fig8}(b).

The mean absolute error ($Err1$), the mean absolute error percentage ($Err2$) and the standard deviation ($SD$) are defined to further evaluate the accuracy of the proposed AM-GAN by performing HR and RR calculations. Table \ref{Tab1} presents the $Err1$ and $Err2$ values for the cycling and treadmill exercise. The average across the 16 testing subject datasets, $\overline{Err1}$ of the proposed AM-GAN is $1.37 \pm 0.32 $ beats/min (Mean $\pm$ SD), and that of $\overline{Err2}$ is $1.29\% \pm 0.35\%$ for the HR calculation. $\overline{Err1}$ is $2.49 \pm 0.52$ breaths/min (Mean $\pm$ SD) and $\overline{Err2}$ is $10.78\% \pm 1.94\%$ for the RR calculation.

\begin{table}[!hbt]
\renewcommand{\arraystretch}{1.5}
\centering
\caption{Mean Absolute Error ($Err1$) (in beats/min) and Average Error Percentage ($Err2$) (\%) for all subjects (F-female, M-male, T-treadmill, C-cycling) of the in-house LU dataset during the cycling and treadmill exercises.}
\label{Tab1}
\resizebox{1\textwidth}{!}{
\begin{tabular}{p{0.65cm} p{0.65cm} p{0.65cm} p{0.65cm} p{0.65cm} p{0.65cm} p{0.65cm} p{0.65cm} p{0.65cm} p{0.65cm} p{0.65cm} p{0.65cm} p{0.65cm} p{0.65cm} p{0.65cm} p{0.65cm} p{0.65cm} p{0.65cm}}
\toprule
	& Subj & F01T& M04& M07& M10&   F01C	&F02 &F03&	M04&	M05	&M06&	M07&	M08&	M09&	M10&	M11&	M12\\
\hline
HR	& $Err1$ &	1.25&	1.61&	1.61&	1.99&	1.44&	0.94&	1.10&	1.96&	1.09&	0.99&	1.37&	1.34& 1.28&  1.38&1.30& 0.99\\
&$Err2$&	0.94&	1.26&	1.43&	1.85&	1.22&	0.71&	0.96&	1.83&	1.13&	0.82&	1.44&	1.73& 1.22&  1.59&1.23& 0.93\\
\hline
RR	&$Err1$&	2.67&	2.83&	3.15&	3.43&	2.75&	2.94&	2.19&	2.32&	1.55&	3.10&	1.96&	2.50& 2.21& 2.19&  2.06& 1.93\\
&$Err2$&	7.31&	7.54&	11.66&	12.83&	9.06&	9.75&	12.10&	11.04&	10.46&	13.43&	10.16&	14.55& 9.90& 11.03& 10.28& 11.43\\
\bottomrule
\end{tabular}
}
\end{table}

\textbf{HR calculation on IEEE-SPC dataset:} The performance of AM-GAN was assessed using the IEEE SPC dataset as a benchmark, which includes 12 training datasets and 10 testing datasets, detailed in Table \ref{Tab2}. $Err1$, $Err2$, and $SD$ were calculated and compared with results from other methods using a testing dataset of 10 subjects. This dataset included data collected under strenuous exercise conditions such as weight-lifting, swimming, boxing, and handshaking. In these scenarios, the prominence of MAs can make the assessment of physiological parameters more difficult. This table shows that, for 10 test subjects with the most noise, the AM-GAN produced HR calculations with $Err1$ of 1.81 and $SD$ of 0.76 beats/min, and also with $Err2$ of 1.86 and $SD$ of 1.63 \%. From the $Err1$ and $Err2$ values, the AM-GAN has a better performance compared to the other traditional de-noised methods, but it is not yet as accurate as the CorNET \cite{b9biswas2019cornet} network that adopts a deep learning architecture. CorNET and AM-GAN use different forms of output; the input of CorNET is the PPG signal and the output is the HR value, which is more of a black-box structure and does not clearly reflect the characteristics of the signals corresponding to different HRs. On the contrary, the AM-GAN has an input of PPG signal with noise and an output of noise-free PPG signal, which is used to calculate HR as well as other physiological parameters using the noise-free PPG signal. Additionally, the difference in $Err1$ between the proposed AM-GAN and the reference method (MR) suggests a fundamental similarity in their performance. Meanwhile, the $SD$ of the proposed method ($SD=0.76$  beats/min) is smaller than that of the MR method ($SD=1.42$  beats/min), it signifies a reduced intergroup error value for the proposed AM-GAN, indicative of superior robustness. Furthermore, the proposed AM-GAN incorporates an Attention Mechanism ($Err1=1.81$  beats/min), enhancing the precision of physiological parameter calculation, in contrast to our previous PPG-GAN method without utilising such a mechanism ($Err1=1.93$  beats/min). 

\begin{table}[!hbt]
\renewcommand{\arraystretch}{2}
\centering
\caption{$Err1(Err2)$ (in beats/min and \%) values for the AM-GAN method compared to some other recently reported results, obtained on the IEEE-SPC dataset.}
\label{Tab2}
\resizebox{1\textwidth}{!}{
\begin{tabular}{
 >{\centering\arraybackslash}p{1cm}
>{\centering\arraybackslash}p{1.1cm}
>{\centering\arraybackslash}p{1.7cm}
>{\centering\arraybackslash}p{1.5cm}
>{\centering\arraybackslash}p{1.7cm}
>{\centering\arraybackslash}p{1.7cm}
>{\centering\arraybackslash}p{1.7cm}
>{\centering\arraybackslash}p{1.7cm}
>{\centering\arraybackslash}p{1.7cm}
>{\centering\arraybackslash}p{1.5cm}
>{\centering\arraybackslash}p{1.7cm}
>{\centering\arraybackslash}p{1.5cm}
>{\centering\arraybackslash}p{1.5cm}
>{\centering\arraybackslash}p{1.5cm}
>{\centering\arraybackslash}p{1.5cm}
}
\toprule
Subject	& Activity& TROIKA \cite{b6zhang2014troika} & JOSS \cite{b25zhang2015photoplethysmography}  & SPECMAR \cite{b26islam2019specmar}   & MURAD \cite{b27chowdhury2016real}  &  Temko \cite{b28temko2017accurate} & ANFA \cite{b18zheng2022adaptive}& MR method \cite{b21zheng2023rapid}& CorNET \cite{b9biswas2019cornet} & PPGANET \cite{b33sawangjai2025removal} & MAICR \cite{b34hao2024ppg}& EPDA \cite{b36bradley2024opening}& PPG-GAN \cite{b29zheng2022ppg}& Proposed AM-GAN\\
\hline

1& T2& 6.63(8.76)& 8.07(10.90)&   6.57&   6.35(7.99)&     9.59(10.90)&    4.99(7.52)&  5.69(7.63)& 1.60& 3.79& 4.36& 6.41& 3.67& 3.46(6.29)\\
2& T2& 1.94(2.56)& 1.61(2.01)&   1.76&   1.15(1.58)&    3.65(2.43)&   1.46(1.99)&  1.02(1.35)&  0.24& 1.47& 1.06& 1.82& 1.35&   1.21(1.45)\\
3& T3& 1.35(1.04)& 3.10(2.69)&   2.28&   1.51(1.25)&    3.90(1.51)&   1.77(0.94)&  1.02(0.83)&  1.60& 1.58& 2.57& 2.68& 1.53&   1.46(0.89)\\
4& T3& 7.82(4.88)& 7.01(4.49)&   2.77&   2.96(1.87)&    2.44(1.49)&   2.89(1.51)&  1.89(1.27)&  2.04& 2.83& 1.32& 3.01& 2.75&   2.66(1.53)\\
5& T3& 2.46(2.00)& 2.99(2.52)&   2.94&   2.78(2.24)&    2.14(1.70)&   2.93(0.93)&  1.83(1.55)&  0.95& 2.42& 1.89& 2.52& 2.25&   2.09(1.81)\\
6& T3& 1.73(1.27)& 1.67(1.23)&   4.80&   1.54(1.12)&    2.60(0.90)&   1.98(1.01)&  1.15(0.85)&  0.28& 1.69& 1.66& 1.79& 1.44&   1.31(0.99)\\
7& T2& 3.33(3.90)&	2.80(3.46)&	  2.72&	  2.03(2.24)&	1.86(1.78)&	  2.01(3.65)&  1.82(2.10)&  0.28& 1.99& 2.41& 1.98& 1.96&   1.73(2.45)\\
8& T3& 3.41(2.43)&	1.88(1.32)&	  3.28&	  2.02(1.61)&	0.85(1.96)&	  1.92(1.31)&  1.73(1.22)&  0.67& 1.81& 1.84& 2.01& 1.77&   1.68(1.21)\\
9& T3& 2.69(2.12)&	0.92(0.74)&	  1.55&	  0.97(0.76)&	3.06(0.80)&	  1.86(1.04)&  0.99(0.80)&  0.42& 1.86& 2.63& 2.22& 1.58&   1.52(1.13)\\
10& T2& 0.51(0.59)&	0.49(0.57)&	  0.82&	  0.83(0.66)&	3.38(0.59)&	  1.16(0.97)&  0.84(0.98)&  0.57& 1.03& 0.81& 1.36& 0.99&   0.93(0.90)\\
 
\hline

Mean&	1-10& 3.19(2.96)&	3.05(2.99)&	2.95&	2.21(2.13)&		3.35(2.41)&	       2.30(2.09)&    1.80(1.85)& 0.86& 2.04& 2.05& 2.58& 1.93& 1.81(1.86)\\
SD  &   1-10&   2.32(2.41)   &   2.53(3.04)&   1.67&   1.62(2.13)&    2.37(3.04)&  1.99(2.08)&  1.42(2.06)& 0.65& 0.79& 1.01& 1.43& 0.79& 0.76(1.63)\\

\bottomrule
\end{tabular}
}
\end{table}

\textbf{Cross-dataset HR validation:} Generalisation capacity is crucial for real-time physiological monitoring. To validate this aspect, cross-database evaluations have been performed to assess the generalisation ability of the proposed AM-GAN. The AM-GAN is trained on the in-house LU dataset and tested on the PPGDalia dataset for HR calculation. The proposed AM-GAN is compared with our previous adversarial network model without an attention mechanism, i.e., PPG-GAN \cite{b29zheng2022ppg}. The cross-dataset HR calculation results of the proposed AM-GAN and the PPG-GAN are demonstrated in Table \ref{Tab3}.

\begin{table}[!hbt]
\renewcommand{\arraystretch}{2}
\centering
\caption{$Err1$ values for cross-dataset HR calculation (training on LU dataset and testing on PPGDalia dataset) by the proposed AM-GAN and PPG-GAN method to compare with the SOTA methods. p-values for a two-sided Wilcoxon signed-rank test against the AM-GAN model.}
\label{Tab3}
\resizebox{1\textwidth}{!}{
\begin{tabular}{
 >{\centering\arraybackslash}p{3.5cm}
>{\centering\arraybackslash}p{0.7cm}
>{\centering\arraybackslash}p{0.7cm}
>{\centering\arraybackslash}p{0.7cm}
>{\centering\arraybackslash}p{0.7cm}
>{\centering\arraybackslash}p{0.7cm}
>{\centering\arraybackslash}p{0.7cm}
>{\centering\arraybackslash}p{0.7cm}
>{\centering\arraybackslash}p{0.7cm}
>{\centering\arraybackslash}p{0.7cm}
>{\centering\arraybackslash}p{0.7cm}
>{\centering\arraybackslash}p{0.7cm}
>{\centering\arraybackslash}p{0.7cm}
>{\centering\arraybackslash}p{0.7cm}
>{\centering\arraybackslash}p{0.7cm}
>{\centering\arraybackslash}p{0.7cm}
>{\centering\arraybackslash}p{2.5cm}
>{\centering\arraybackslash}p{1.5cm}
}
\toprule
	& S1& S2 & S3& S4&  S5&  S6& S7& S8& S9& S10& S11& S12& S13& S14& S15& Mean (SD)& p-value\\
\hline
DeepPPG \cite{b29reiss2019deep}& 7.73& 6.74& 4.03& 5.90& 18.51& 12.88&	3.91&	10.87&	8.79&	4.03&	9.22&	9.35&	4.29&	4.37&	4.17& 7.65 (4.15)& \textbf{6.08e-5}\\
NAS-PPG \cite{b30risso2021robust} & 5.46&	5.01&	3.74&	6.48&	12.68&	10.52&	3.31&	8.07&	7.91&	3.29&	7.05&	6.76&	3.84&	4.85&	3.57& 6.17 (2.77)& \textbf{6.10e-5}\\
Q-PPG \cite{b31burrello2021q} & 4.29&	3.62&	2.44&	5.73&	10.33&	5.26&	2.00&	7.09&	8.60&	3.09&	4.99&	6.25&	1.92&	3.02&	3.55& 4.81 (2.46)& \textbf{1.53e-3}\\
Aug-TEMPONet \cite{b32burrello2022improving}&  4.97&	4.34&	2.39&	6.14&	9.41&	3.63&	2.23&	9.14&	10.98&	3.40&	5.27&	7.64&	2.05&	2.84&	3.61& 5.20 (2.86)& \textbf{8.54e-4}\\
PPGANET \cite{b33sawangjai2025removal} & 4.51&	3.37&	2.33&	4.25&	8.68&	4.76&	2.37&	7.04&	7.75&	2.80&	5.19&	6.08&	2.09&	3.02&	3.19&   4.49 (2.08)& \textbf{2.01e-3}\\
Adaptive Q-PPG \cite{b38kechris2024kid} & 3.80&	3.50&	2.07&	5.18&	\textbf{5.76}&	\textbf{3.38}&	\textbf{1.59}&	\textbf{6.99}&	8.94&	2.75&	3.37&	\textbf{5.61}&	\textbf{1.49}&	\textbf{2.45}&	2.80&   3.98 (2.12)& 7.76e-1 \\
PULSE \cite{b37kasnesis2023feature} & 4.75&	3.31&	2.22&	5.25&	7.43&	4.22&	2.28&	8.93&	6.95&	2.93&	3.98&	6.57&	1.70&	3.22&	2.88&   4.44 (2.16)& 1.03e-2\\
PPG-GAN \cite{b29zheng2022ppg}& 4.81&	3.85&	2.62&	5.31&	9.63&	4.69&	2.32&	9.99&	8.06&	3.54&	3.93&	6.68&	1.84&	3.42&	2.95&   4.91 (2.57)& \textbf{1.83e-4}\\
AM-GAN & \textbf{3.01}&	\textbf{2.98}&	\textbf{2.04}&	\textbf{3.97}&	7.90&	4.64&	2.41&	7.12&	\textbf{5.83}&	\textbf{2.33}&	\textbf{3.09}&	6.38&	1.73&	2.49&	\textbf{2.01}&   \textbf{3.86 (2.02)}& \\

\bottomrule
\end{tabular}
}
\end{table}

The comprehensive cross-dataset HR calculation outcomes for both the proposed AM-GAN and the PPG-GAN are presented in Table \ref{Tab3}. The proposed AM-GAN illustrated the best results, given mean HR calculations with $\overline{Err1}$ of $3.86$ and $SD$ of $2.02$ beats/min, compared with PPG-GAN and SOTA methods. However, subjects S5, S6 and S8 have posed challenges with high $Err1$ values. Comparison of S5, S6 and S8 shows that the underlying causes of high errors are different. S5 and S6 are primarily affected by the out-of-distribution samples, and S8 is mainly affected by the lower signal quality of collected data. Pairwise comparisons using the two-sided Wilcoxon signed-rank test with Bonferroni correction indicated that the proposed AM-GAN method significantly outperforms DeepPPG, NAS-PPG, Q-PPG, Aug-TEMPONet, PPGANET, and PPG-GAN ($p < 0.05$). No statistically significant differences were found between AM-GAN and Adaptive Q-PPG or PULSE. In summary, the cross-dataset HR validations indicate that the proposed AM-GAN exhibits robust generalisation capabilities in scenarios characterised by unknown noise and MAs.

\begin{table}[!hbt]
\renewcommand{\arraystretch}{1.5}
\centering
\caption{Comparison of $Err1(Err2)$ (in \% SpO$_2$ and \% error) for cross-dataset SpO$_2$ estimation. Training was done on LU dataset and testing on C2 dataset under cycling (C) condition.}
\resizebox{\textwidth}{!}{
\begin{tabular}{lcccccccccccccc}
\toprule
Subject & S1 & S1 & S1 & S2 & S2 & S2 & S3 & S3 & S3 & S4 & S4 & S4 & \multicolumn{1}{c}{\textbf{Mean $\pm$ SD}}  \\
Oxygen Level & 21\% & 18\% & 16\% & 21\% & 18\% & 16\% & 21\% & 18\% & 16\% & 21\% & 18\% & 16\% &  & \\
\hline
PPG-GAN\cite{b29zheng2022ppg} (C) & 
2.39 & 2.78 & 4.08 &
2.02 & 2.68 & 1.68 &
1.17 & 1.53 & 1.47 &
1.97 & 1.55 & 2.74 & 2.17$\pm$0.78 \\
 & 
(2.51) & (2.99) & (4.90) &
(2.03) & (2.88) & (1.80) &
(1.20) & (1.60) & (1.55) &
(2.00) & (1.59) & (2.97) & (2.33$\pm$0.96)  \\
AM-GAN (C) & 
1.28 & 2.04 & 3.93 &
0.80 & 2.18 & 1.35 &
1.49 & 1.37 & 1.09 &
0.96 & 1.95 & 1.47 & \textbf{1.65$\pm$0.79} \\
 & 
(1.35) & (2.19) & (4.73) &
(0.81) & (2.36) & (1.45) &
(1.52) & (1.42) & (1.14) &
(0.97) & (1.99) & (1.59) & \textbf{(1.79$\pm$0.99)}   \\
\bottomrule
\end{tabular}
}
\label{Tab5}
\end{table}

\textbf{Cross-dataset SpO$_2$ validation:} For an in-depth exploration of the AM-GAN's robustness, cross-dataset SpO$_2$ validation results are outlined in Table \ref{Tab5}. Notably, given that the in-house C2 dataset incorporates a multi-wavelength opto-physiological monitoring sensor, the cross-dataset SpO$_2$ validation necessitates the utilisation of two additional wavelength channels, i.e., red and infrared channels. The relationship between SpO$_{2}$ and the ratio of absorbance $\rho$ \cite{b21zheng2023rapid} is provided by the C2 experiments. 

\begin{equation}
\label{e11}
\textrm{SpO}_2 \% = 109-21.5\rho
\end{equation}
where $\rho=(AC_{Red}/DC_{Red})/(AC_{IR}/DC_{IR} )$. 

\begin{figure}[!hbt]
    \centering
    \includegraphics[width=1\textwidth]{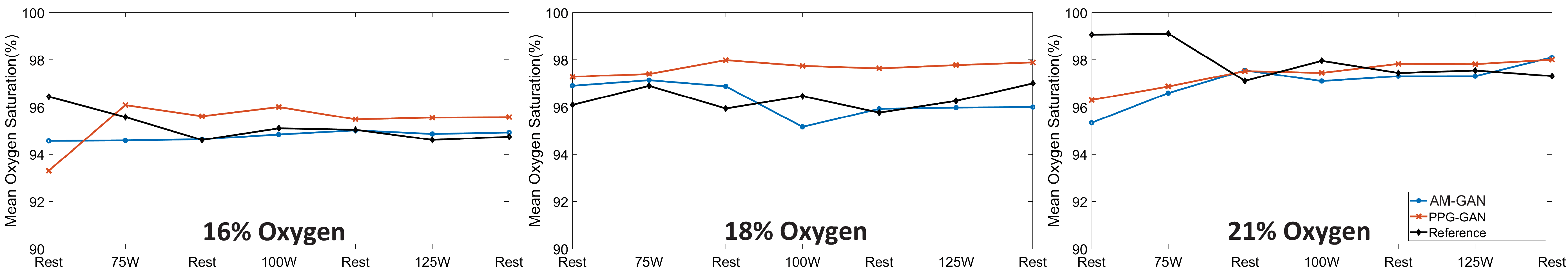}
    \caption{SpO$_2$ calculation results from C2 subject 4 (three oxygen levels) in cycling exercise. The X-axis illustrates the cycling activity intensities.}
    \label{Fig_spo2}
\end{figure}

From Table \ref{Tab5}, the AM-GAN performance ($\overline{Err1}(\overline{Err2})$ of $1.65(1.79)$ and $SD$ of $0.79(0.99)$\%) consistently outperforms our prior approach, which lacked an attention mechanism, across all evaluation metrics. Fig. \ref{Fig_spo2} demonstrated that the calculated SpO$_2$ from AM-GAN closely aligns with the reference SpO$_2$ in three different oxygen levels. The performance of SpO$_{2}$ calculation exhibits commendable alignment with the reference SpO$_{2}$, revealing the $Err1$ of $1.65$\% for cycling exercises in three different oxygen levels. The cross-dataset SpO$_2$ validations indicate that the proposed framework exhibits robust generalisation capabilities in scenarios characterised by unknown noise and motion artefacts.

\textbf{Ablation study:} The ablation study is performed to evaluate the performance of AM-GAN using the in-house LU dataset. The following ablation study is covered: (I) without triaxial ACC as input; (II) without discriminator; (III) without attention mechanism; and (IV) the whole AM-GAN method. The results are presented in Table \ref{Tab4}. 

\begin{table}[!hbt]
\normalsize
\renewcommand{\arraystretch}{1.5}
\centering
\caption{Ablation study of proposed AM-GAN in terms of attention mechanism, discriminator and triaxial ACC input for HR and RR validation on in-house LU dataset and SpO$_2$ on in-house C2 dataset.}
\label{Tab4}
\resizebox{1\textwidth}{!}{
\begin{tabular}{
 >{\centering\arraybackslash}p{3.5cm}
>{\centering\arraybackslash}p{3cm}
>{\centering\arraybackslash}p{3cm}
>{\centering\arraybackslash}p{2cm}
>{\centering\arraybackslash}p{2cm}
>{\centering\arraybackslash}p{2cm}
>{\centering\arraybackslash}p{2cm}
>{\centering\arraybackslash}p{2cm}
>{\centering\arraybackslash}p{2cm}
}
\toprule
	Attention Mechanism& Discriminator& triaxial ACC Input & $\overline{Err1}$(HR)& $\overline{Err2}$(HR)&  $\overline{Err1}$(RR)&  $\overline{Err2}$(RR)& $\overline{Err1}$(SpO$_2$)&  $\overline{Err2}$(SpO$_2$)\\
\hline
$\usym{2717}$ & $\usym{2717}$& $\usym{2717}$& 3.81&   3.90&   4.99&   14.86& 3.67& 3.58\\
$\usym{2717}$ & $\usym{2713}$&  $\usym{2717}$&	3.22&	3.69&	4.63&	14.43& 3.11& 3.21\\
$\usym{2717}$ & $\usym{2717}$& $\usym{2713}$&	2.99&	3.03&	4.11&	13.98& 2.86& 2.99\\
$\usym{2717}$ & $\usym{2713}$&  $\usym{2713}$&	2.45&	2.71&	3.65&	13.12& 2.73& 2.80\\
$\usym{2713}$ & $\usym{2717}$&  $\usym{2713}$&	1.89&	1.73&	3.01&	12.35& 2.05& 1.97\\
$\usym{2713}$ & $\usym{2713}$&  $\usym{2713}$&	\textbf{1.37}&	\textbf{1.29}&	\textbf{2.49}&	\textbf{10.78}& \textbf{1.65}& \textbf{1.79}\\

\bottomrule

\end{tabular}
}
\end{table}

Comparing the results of (I) and (II), we can observe that using triaxial ACC input can reduce the $\overline{Err1}$($\overline{Err2}$) by 0.23(0.66) beats/min(\%) for HR validation, 0.52(0.45) breaths/min(\%) for RR and 0.25(0.22) \% for SpO$_2$. This suggests that the triaxial ACC input as multi-sensor fusion is helpful for improving HR, RR, and SpO$_2$ validation accuracy. Similarly, using the attention mechanism for multi-sensor fusion can greatly improve HR, RR, and SpO$_2$ validation accuracy. These results indicate that the attention mechanism integrates inputs from multiple sensors for multi-sensor feature fusion to learn a good distribution of the MAs mixed with the PPG signals. It leads to better PPG feature disentanglement.

\begin{figure}[!hbt]
    \centering
    \begin{minipage}[b]{1\textwidth}
    \subfloat[]{
        \includegraphics[width=0.48\textwidth]{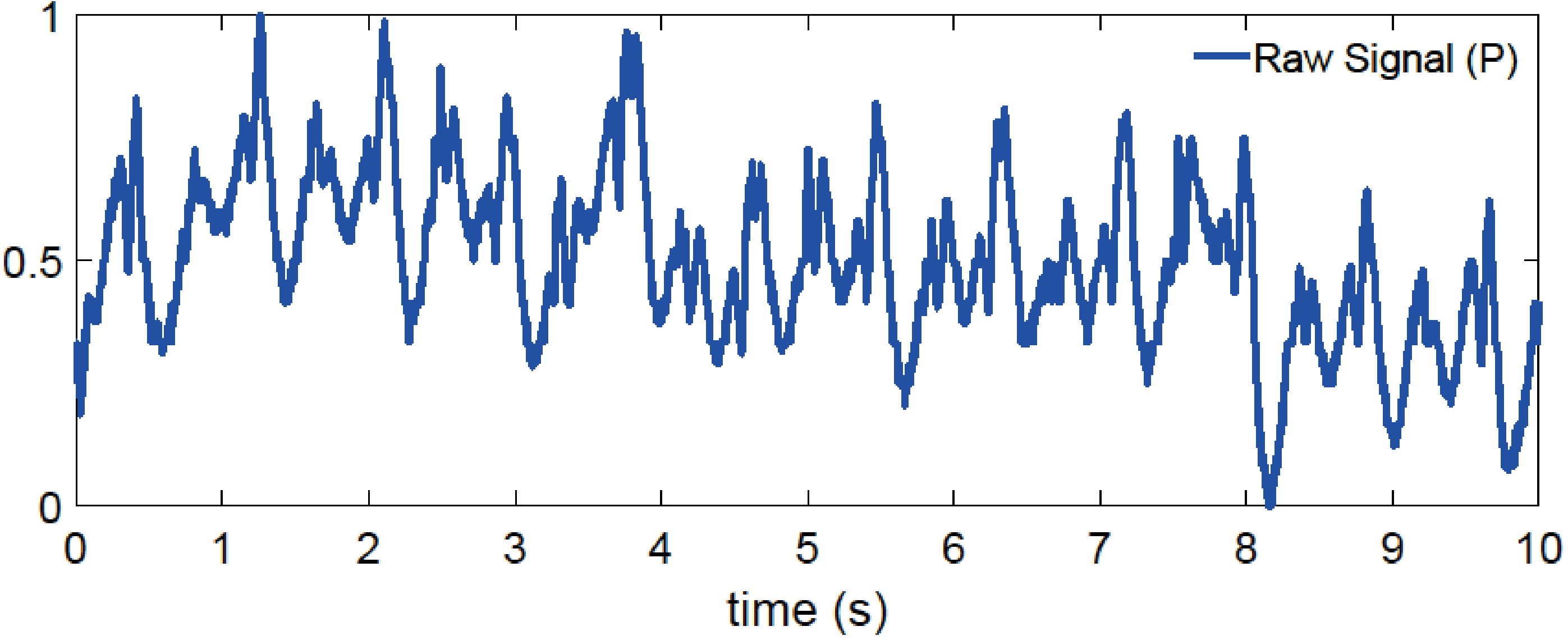}
    }\hspace{0mm}
    \subfloat[]{
        \includegraphics[width=0.48\textwidth]{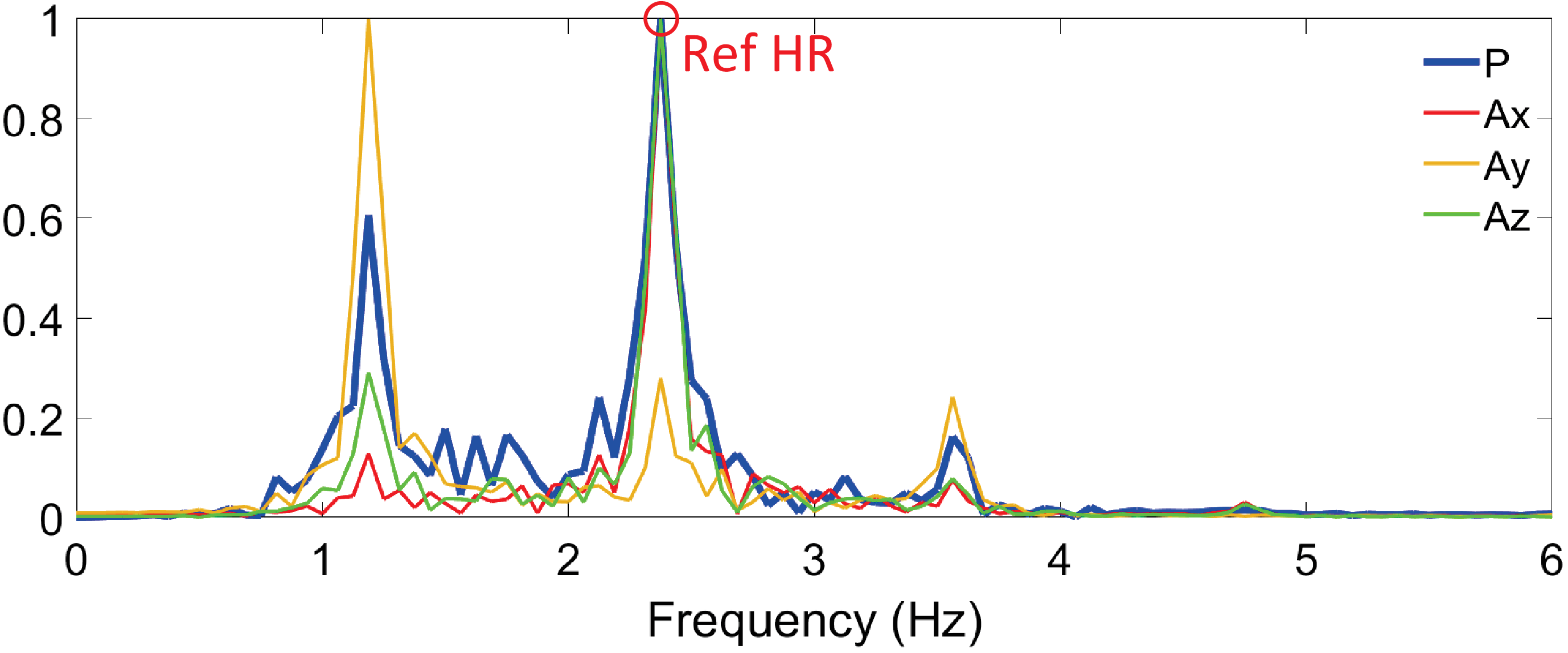}
    }
    \vspace{0mm} 
    \subfloat[]{
        \includegraphics[width=0.48\textwidth]{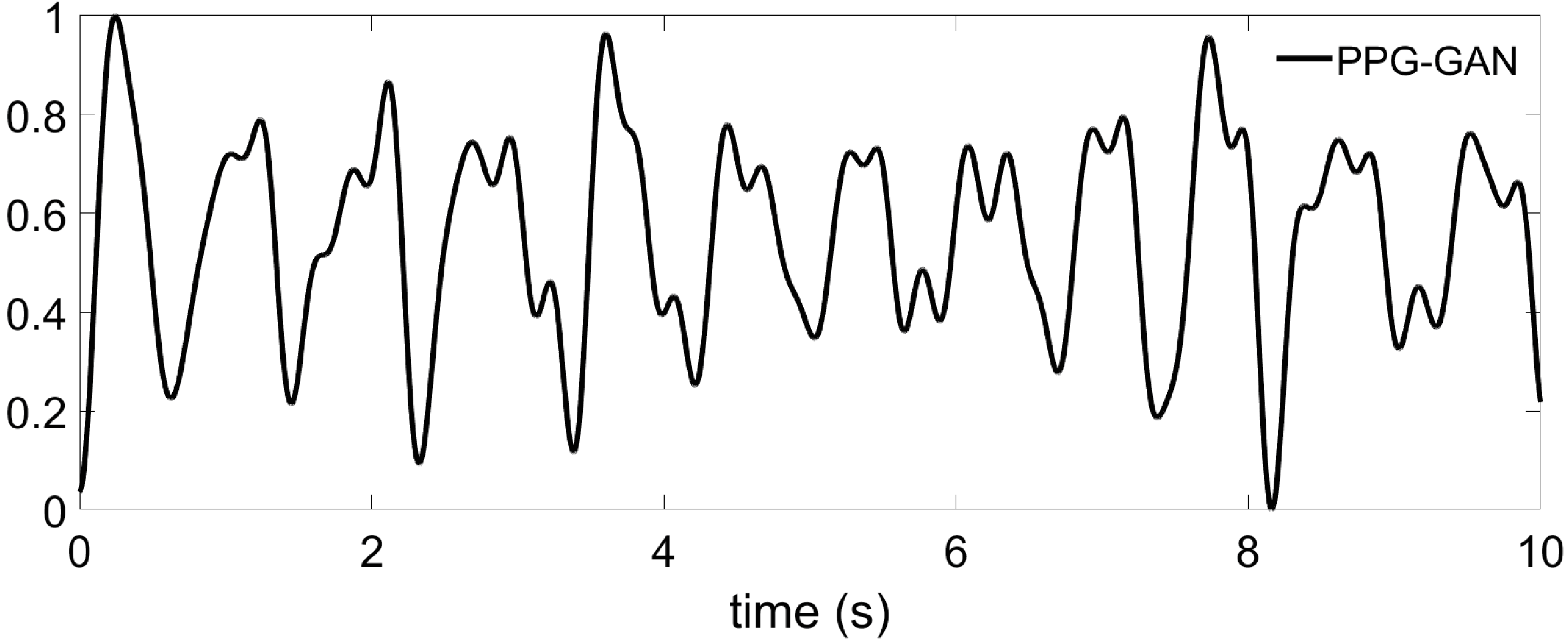}
    }\hspace{-4mm}
    \subfloat[]{
        \includegraphics[width=0.48\textwidth]{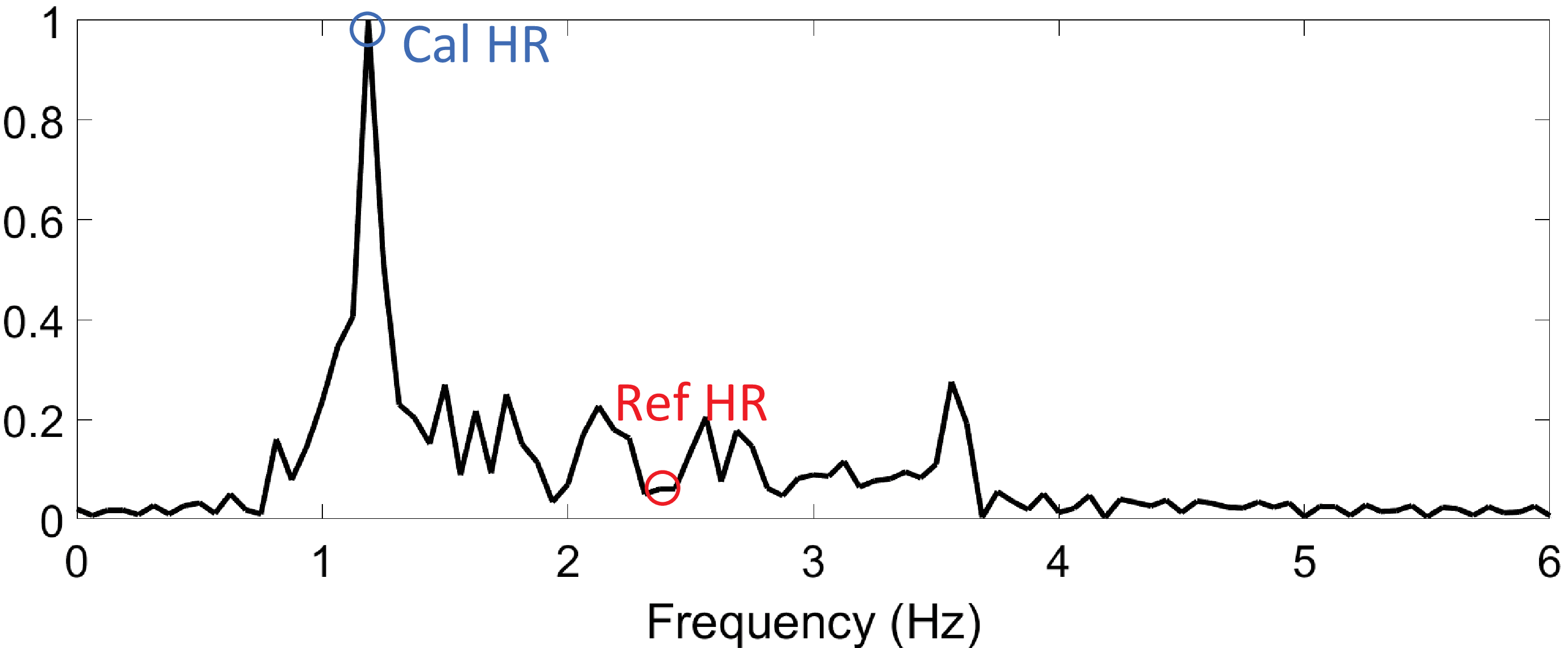}
    }
    \vspace{0mm} 
    \subfloat[]{
        \includegraphics[width=0.48\textwidth]{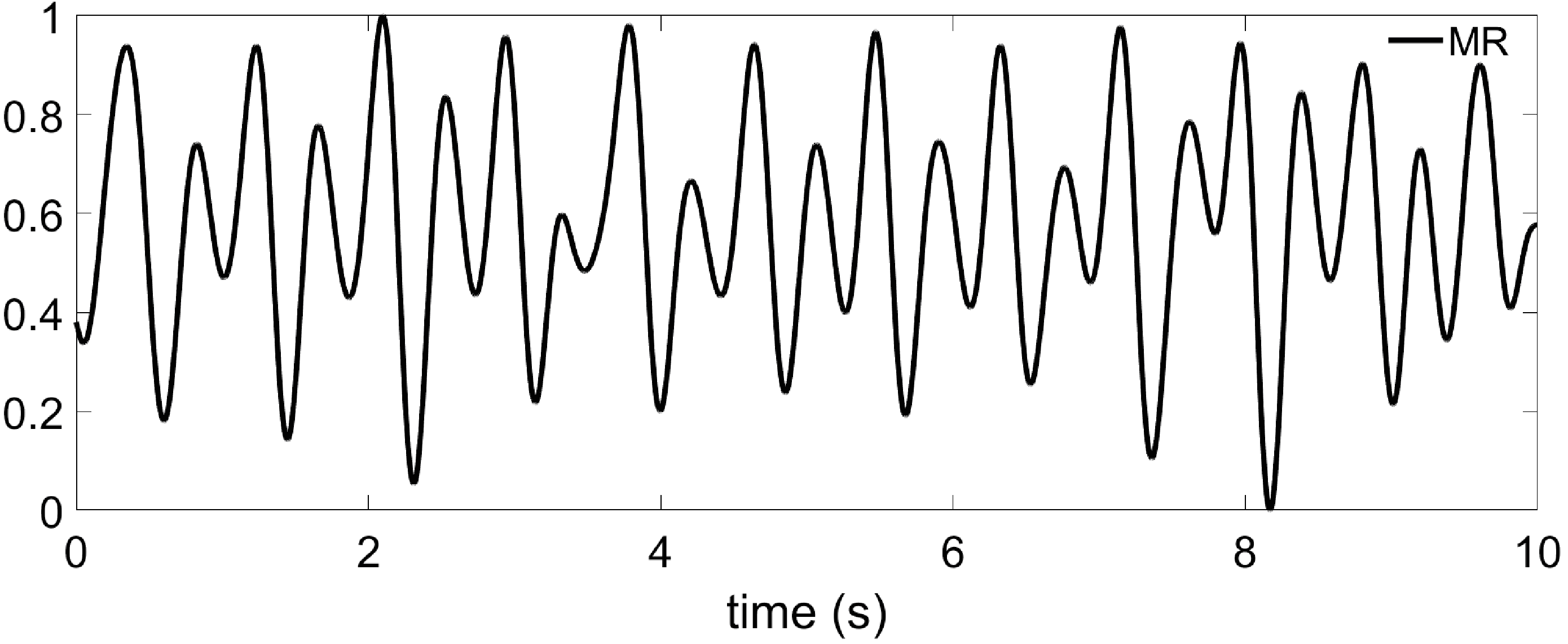}
    }\hspace{0mm}
    \subfloat[]{
        \includegraphics[width=0.48\textwidth]{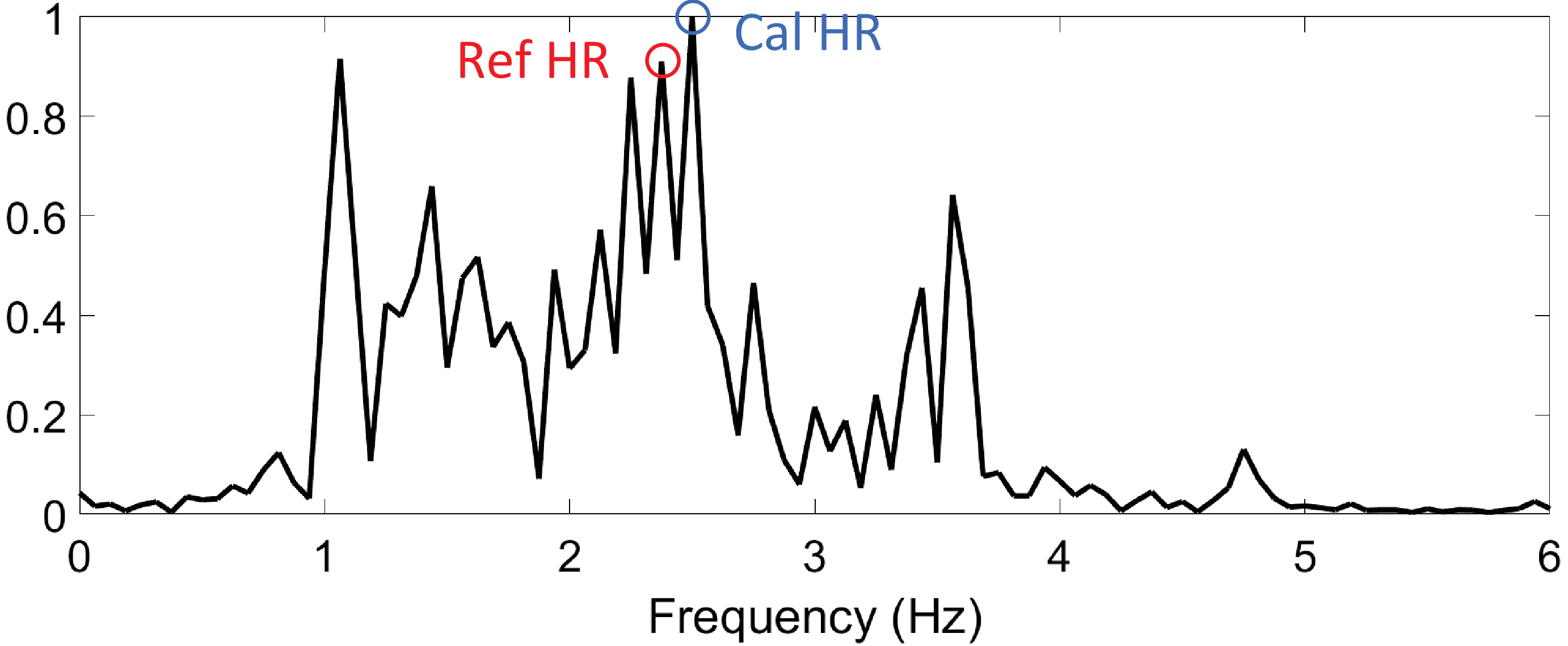}
    }
    \vspace{0mm} 
    \subfloat[]{
        \includegraphics[width=0.48\textwidth]{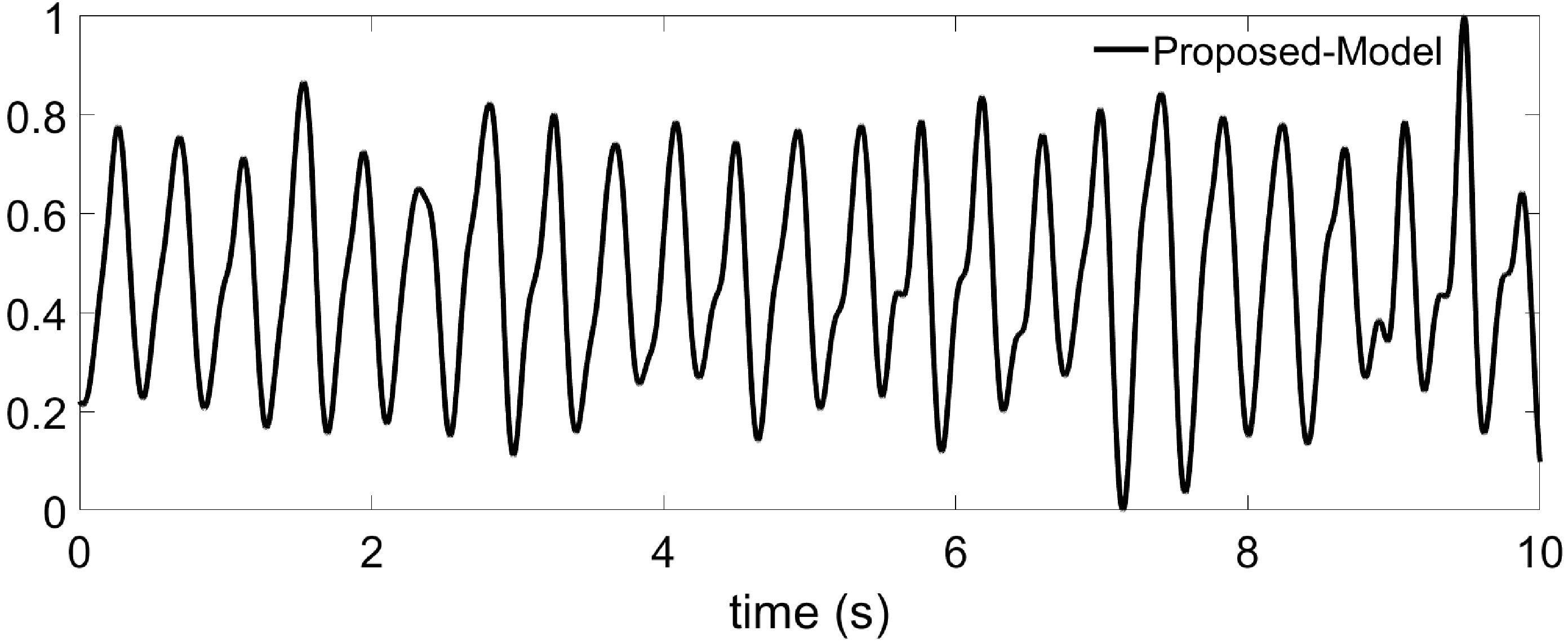}
    }\hspace{0mm}
    \subfloat[]{
        \includegraphics[width=0.48\textwidth]{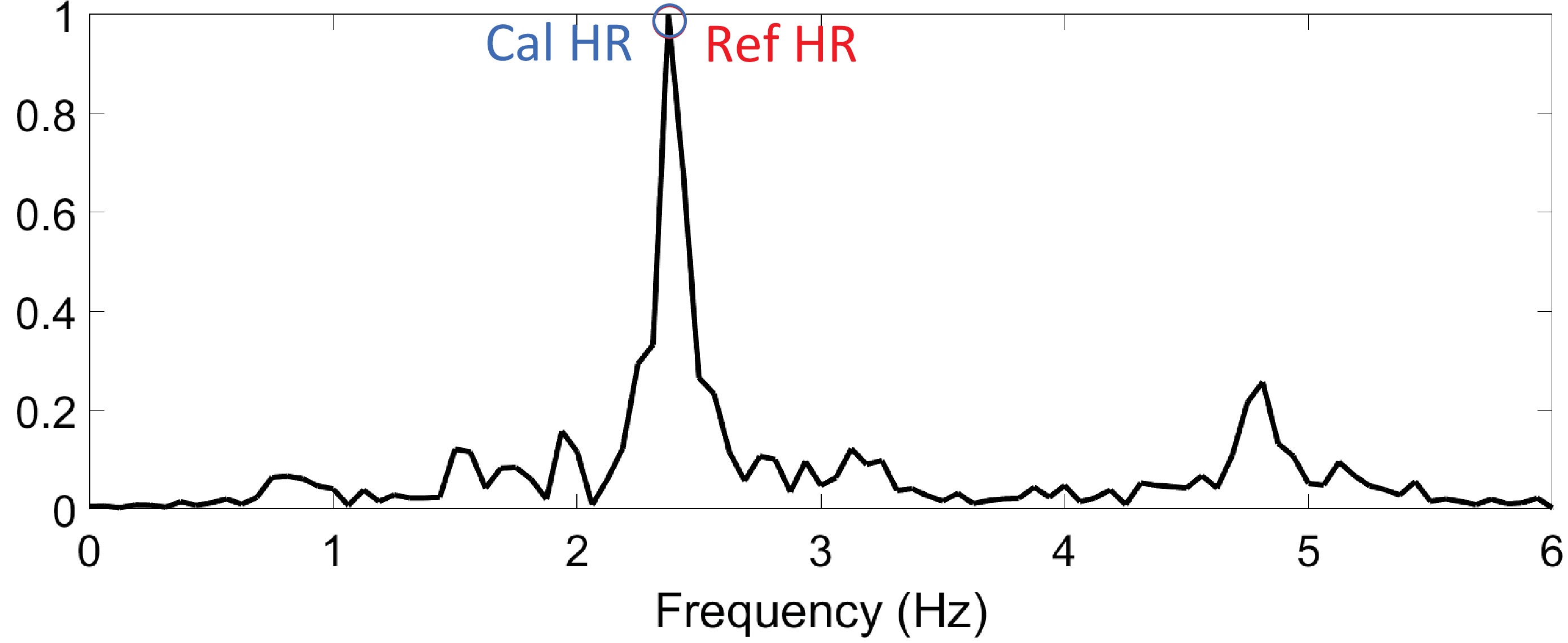}
    }
    \caption{Raw signal and de-noised PPG signals in time and frequency domains. (a) Time-domain raw PPG signal. (b) Frequency-domain raw PPG signal (P) and triaxial ACC signals (Ax, Ay and Az). (c,d) Time and frequency-domain generated de-noised PPG signal from PPG-GAN \cite{b29zheng2022ppg}. (e,f) Time and frequency-domain de-noised PPG signal from MR \cite{b21zheng2023rapid}. (g,h) Time and frequency-domain generated de-noised PPG signal from AM-GAN.}
    \label{Fig11}
    \end{minipage}
    \vspace{-6mm}
\end{figure}

\begin{figure}[!hbt]
    \setlength{\belowcaptionskip}{-0.3cm} 
    \begin{minipage}[b]{1\textwidth}
    \centering
    \subfloat[]{
        \includegraphics[width=0.48\textwidth]{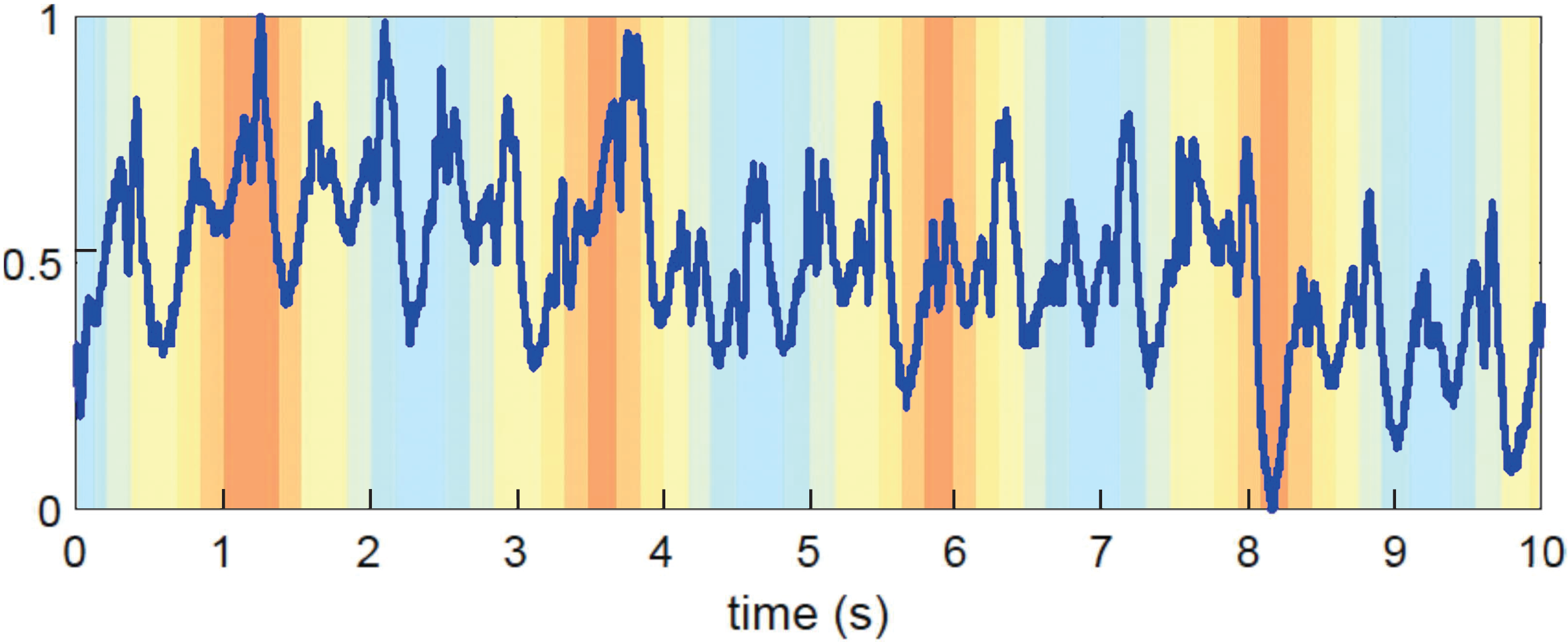}
    }
    \hspace{-3mm} 
    \subfloat[]{
        \includegraphics[width=0.48\textwidth]{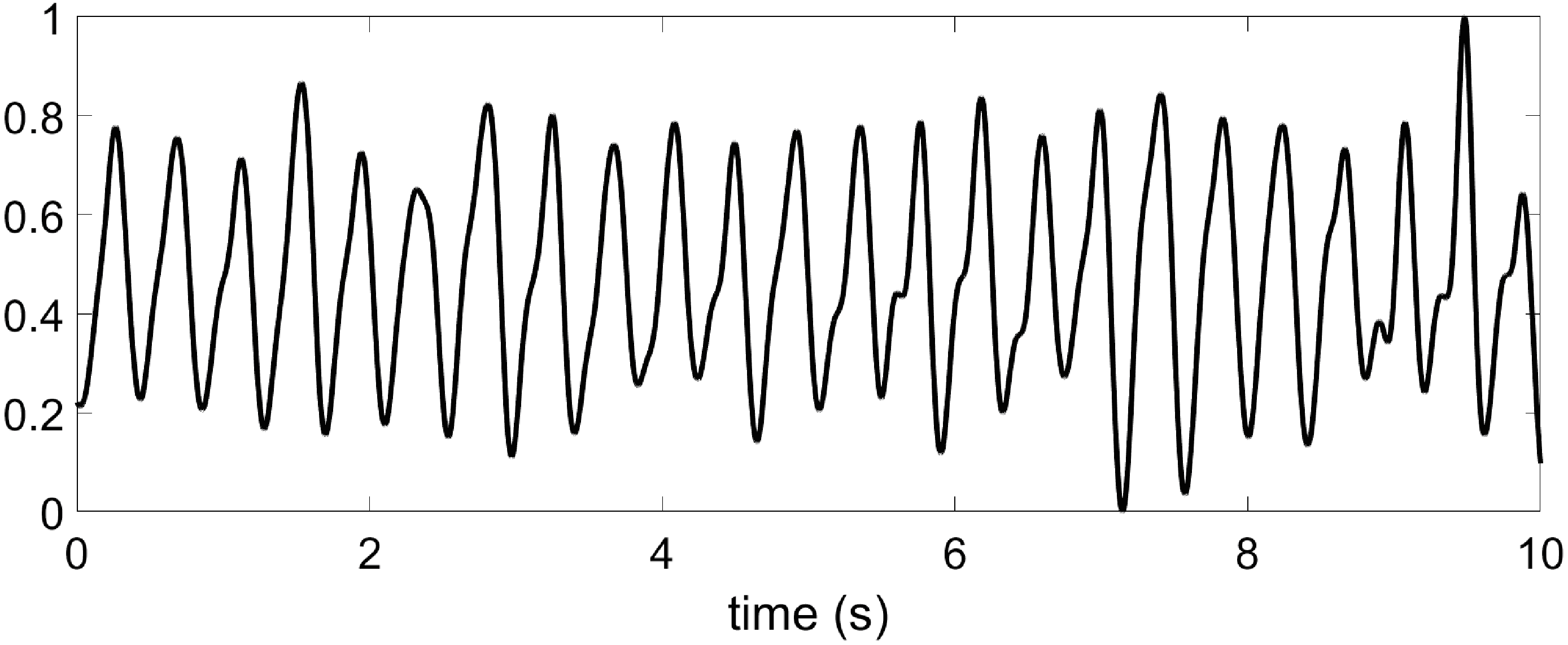}
    }
    \caption{Visualisation of attention maps is presented where the brighter parts indicate regions to which the generator pays more attention. (a) The attention map overlays highlight the raw PPG signal segments. (b) Reconstructed PPG signal from AM-GAN.}
    \label{Fig12}
    \end{minipage}
\end{figure}

Additionally, we present a visualisation of the synthetic outcomes by comparing results (III) and (IV). The generated de-noised PPG signals are demonstrated in time and frequency domains by the without attention mechanism method (PPG-GAN), the SOTA MA-removal algorithm (MR) and the proposed AM-GAN in Fig. \ref{Fig11}. This demonstration presents a complex example from the MW dataset in which the period of the HR and the motion period detected by the three-axis accelerometer sensor align closely, as depicted in Fig. \ref{Fig11}(b). The MR method, which employs triaxial ACC as the motion reference, inadvertently filters out the real HR frequency owing to the exact overlap between the motion reference period and the HR period, as illustrated in Fig. \ref{Fig11}(e) and (f). PPG-GAN is not effective in dealing with this scenario, as shown in Fig. \ref{Fig11}(c) and (d). This also explains why the AM-GAN is able to work as a noise-reduced model across varying intensities of exercise to enhance the robustness in calculating physiological parameters, i.e., HR and RR. As presented in Fig. \ref{Fig12}, the attention heatmap provides interpretability regarding which signal segments contribute the most to the MA-removal process. The attention weights learned by the model specifically reflect the relative importance assigned to different segments of the input raw PPG signal. Also, the red colouration in Fig. \ref{Fig12}(a) denotes regions where the model places higher attention, particularly focusing on MA segments. To evaluate the sensitivity of the PPG-GAN to the choice of optimizer and learning rate (LR), we also conducted an ablation study (Table \ref{tab:optimizer-lr}) in which the model was trained using three widely adopted optimizer, i.e., SGD (with momentum set to 0.9), AdamW, and Adam, each tested at three different learning rates: 0.01, 0.001, and 0.0002. All experiments were performed using the IEEE-SPC dataset.


\begin{table}[h]
\small
\centering
\caption{Comparison of different optimizers and learning rates.}
\label{tab:optimizer-lr}
\begin{tabular}{lcc}
\toprule
\textbf{Optimizer} & \textbf{LR} & \textbf{$Err_1$ (Mean$\pm$SD \%)} \\
\midrule
SGD         & 0.01   & 2.74$\pm$2.45 \\
SGD         & 0.001  & 2.53$\pm$2.46 \\
SGD         & 0.0002 & 2.43$\pm$2.84 \\
AdamW       & 0.01   & 2.22$\pm$2.15 \\
AdamW       & 0.001  & 2.32$\pm$2.69 \\
AdamW       & 0.0002 & 2.01$\pm$2.12 \\
Adam (ours) & 0.01   & 2.09$\pm$1.99 \\
Adam (ours) & 0.001  & 2.06$\pm$2.33 \\
Adam (ours) & 0.0002 & \textbf{1.81$\pm$1.81} \\
\bottomrule
\end{tabular}
\end{table}

\section{Discussion} 
The study has researched and developed an adversarial learning strategy with an attention mechanism to obtain clinically acceptable physiological parameters through PPG signals acquired from those subjects engaged in exercise at varying intensities. In high-intensity scenarios, the measured signals exhibit significant distortion, primarily due to motion artefacts (MAs), which are further exacerbated by the potential overlap between motion frequencies and cardiac waveform frequencies. It is essential to eliminate MAs before evaluating any physiological parameters. The output from the triaxial accelerometer to act as motion reference signals was acquired along with four-channel PPG signals of mOEPS, and MAs were removed using the adversarial network with an attention mechanism (AM-GAN). The values for HR,  RR, and SpO$_2$ were extracted from the generated MA-free signal, showing the AM-GAN to have accurate vital signs even in the state of higher-intensity exercise. The analysis of these outcomes across three distinct datasets reveals a substantial enhancement in accuracy attributable to the consolidation of the modified GAN and the multi-head cross-attention mechanism.

The pulsatile waveform, quantified by the Pearson correlation coefficient $R$, is illustrated in Table \ref{Fig13}. Fig. \ref{Fig13}(a) and (b) evidence a high degree of consistency between the generated PPG signals and the reference signal across all LU datasets ($\overline{\langle R \rangle}>90$). The incorporation of the in-house LU dataset has facilitated the acquisition of benchmark readings for HR and RR. This benchmarking is crucial for assessing the accuracy of the AM-GAN method in extracting these vital measurements. The IEEE-SPC dataset and PPGDalia dataset served as a benchmark for comparing the performance of the AM-GAN against SOTA algorithms. 

Table \ref{Tab2} demonstrates that the proposed AM-GAN achieves higher accuracy in HR extraction from the IEEE-SPC dataset compared to SOTA algorithms. The CorNET framework achieves more precise HR extraction (0.86 beats/min \textless 1.81 beats/min) compared to the AM-GAN; however, it is limited to assessing only HR. In contrast, the AM-GAN is designed to deliver a complete pulsatile waveform with MAs removed, enabling the extraction of multiple physiological parameters, i.e., HR, RR and SpO$_2$. The precision of RR extraction demonstrates satisfactory alignment with the Vyntus$^\text{TM}$ CPX Metabolic cart, manifesting a mean absolute error of $3.02$ breaths/min and $2.31$ breaths/min for cycling and treadmill exercises, respectively. The mean absolute error associated with RR measurements exceeds that of HR measurements. This difference can be attributed to the fact that RR frequencies fall within a lower frequency range (0.1-1.0 Hz), rendering them more vulnerable to motion interference compared to HR extraction.   

AM-GAN also has a better performance for cross-dataset (PPGDalia) HR validation with $\overline{Err1}$ of $3.86$ and $SD$ of $2.02$ beats/min compared with the SOTA methods, as shown in Table \ref{Tab3}. It is presented that the AM-GAN has robust generalisation capabilities in scenarios characterised by unknown noise and MAs. However, some of the subject data have posed challenges with high $Err1$ values, such as S5, S6 and S8. A comparison of S5, S6 and S8 demonstrates that the lower signal quality of collected data and out-of-distribution HR values could cause the high $Err1$ values. Therefore, experimental outcomes using the in-house LU dataset, C2 dataset, PPGDalia dataset, and IEEE-SPC dataset indicate that the proposed AM-GAN effectively eliminates MAs and improves the quality of noisy input PPG signals. The resulting clean PPG signal enables the determination of HR, RR and SpO$_2$ with an accuracy exceeding 95\% when compared to gold standard references. Additionally, the ablation study is performed to validate the interpretability of the AM-GAN method, as shown in Table \ref{Tab4}. The visualisation of the generated de-noised PPG signals also presents better performance in the challenge scenarios, i.e., high-intensity exercise.

Although we achieved highly satisfactory results, there is still a lot of room for improvement. First, our work primarily focuses on removing motion artefacts due to the availability of such published datasets. However, other artefacts, such as muscular artefacts, may also need to be addressed in certain scenarios. Second, our approach requires a reference PPG signal from state-of-the-art MA removal algorithms during the training phase. The performance of AM-GAN may be significantly affected if the generated ground truth is of suboptimal quality. While getting high-quality ground truth signals can be challenging, their impact may be mitigated. Employing advanced training techniques, innovative model architectures, and specialized model layers could reduce the dependence on large amounts of training data.

In the current state, the AM-GAN generator has a significant number of parameters, leading to a memory footprint of approximately 30 MB and requiring around 2 GFLOPs per inference. On a standard CPU, this physiological monitoring results (HR, RR and SpO$_2$) in an inference latency of 500ms. For further deployment and optimisation, first, we would explore model pruning to remove redundant weights and channels from the generator. Techniques like magnitude-based pruning could significantly reduce the model's size and computational load. Next, we would apply post-training quantisation. By converting the model's weights from 32-bit floating-point numbers to 8-bit integers (INT8), we can achieve an almost $4\times$ reduction in model size and a significant speed-up in inference. Furthermore, we could employ knowledge distillation. Here, our AM-GAN would act as a 'teacher' model. We would train a much smaller, lightweight 'student' generator to mimic the output of the teacher. Finally, the optimised model would be compiled using a framework like TensorFlow Lite or PyTorch Mobile. These tools are specifically designed to run models efficiently on edge devices, taking advantage of hardware accelerators like Neural Processing Units (NPUs), which are becoming more common in wearable devices.

\section{Conclusion} 
The cardiac signal is critical for computing physiological parameters such as HR, RR, and SpO$_2$. This study introduces the AM-GAN framework, which integrates generative adversarial networks with attention mechanisms (AM-GAN), specifically designed to extract high-quality pulsatile waveforms for enhanced opto-physiological monitoring. The AM-GAN leverages both time- and frequency-domain error losses to effectively enhance noisy PPG signals, using these signals and associated triaxial ACC data as input.

Experimental evaluations on two in-house datasets and one publicly available dataset have demonstrated that AM-GAN consistently improves signal quality, significantly enhancing the accuracy of physiological parameter estimation both within-dataset and across datasets. Comparative analyses with other recent PPG enhancement methods, including $Err1(Err2)$, have confirmed the superior performance of AM-GAN in accurately calculating HR, RR, and SpO$_2$ by producing clearer, more reliable pulsatile waveforms.

Moreover, AM-GAN has shown the ability to learn MA removal algorithms, specifically outperforming the original MR algorithm (MR algorithm, \cite{b21zheng2023rapid}), particularly in scenarios involving complex interactions between cardiac and motion frequencies during intense physical activity. Although the current findings represent preliminary evidence, they clearly indicate substantial potential for future advancements.

Looking ahead, the AM-GAN framework could be expanded to integrate additional physiological signals, further enhancing its robustness and generalizability in various clinical and nonclinical scenarios. Future work could focus on applying this methodology to wearable and remote health monitoring systems, enabling real-time and accurate physiological assessments in environments with high MAs. Additionally, integrating AM-GAN with advanced analytics and deep learning techniques (such as self-supervised learning) could facilitate predictive monitoring, personalised healthcare solutions, and broader telehealth applications, thereby significantly extending the utility and impact of opto-physiological monitoring technologies.

\section*{Author contributions}
XZ carried out the AM-GAN work with the associated comparison study and organized the manuscript. SH supervised, organized, and reviewed the study. VMD extracted multichannel signals and postprocessed the raw datasets. LB supervised the study of physiological monitoring and reviewed the manuscript. MD supervised the study of AM-GAN. 

\section*{Acknowledgments} 
The authors would like to acknowledge the support of Loughborough University in the conduct of the Ph.D. study during the period of October 2020 to December 2024. The authors also acknowledge a) the permit of the C2 dataset for SpO2 calculation from Carelight Limited, UK, b) the implementation of altitude study (Dec 2023 - Jan 2024) to obtain C2 dataset, led by Caroline Sunderland and James Donaldson in Nottingham Trent University, UK, and the shared datasets in the IEEE-SPC and PPGDalia.

\section*{Conflict of interest} The authors declare that there is no any conflict of interest for this study.


\appendix
\counterwithout{table}{section}       
\setcounter{table}{0}                  
\renewcommand{\thetable}{\arabic{table}}
\captionsetup[table]{name=Appendix, labelsep=colon}  
                         
\section*{Appendix.} 
\begin{table}[!htbp]
\raggedright 
\scriptsize
\setlength{\tabcolsep}{5pt}      
\renewcommand{\arraystretch}{1.16}
\caption{Summary of prior MA removal methods.}
\label{tab:lit_review}
\begin{adjustbox}{max width=\textwidth}
\begin{tabular}{%
L{0.16\textwidth}  
L{0.13\textwidth}  
L{0.29\textwidth}  
L{0.14\textwidth}  
L{0.16\textwidth}  
L{0.10\textwidth}  
}
\toprule
\multicolumn{1}{l}{\textbf{Category}} &
\multicolumn{1}{l}{\textbf{Method}} &
\multicolumn{1}{l}{\textbf{Brief Summary}} &
\multicolumn{1}{l}{\textbf{Inputs}} &
\multicolumn{1}{l}{\textbf{Outputs}} &
\multicolumn{1}{l}{\textbf{Exercise Intensity}} \\
\midrule
\multicolumn{6}{l}{\textit{Traditional / Signal-processing approaches}}\\
\midrule
Traditional & EMD/ FFT/ ICA/ Wavelet \cite{b4pankaj2022review} &
Classical decomposition / transform / denoising pipelines for MA reduction. &
PPG &
MA-reduced PPG; HR/ RR/ SpO$_2$ &
Rest--low \\

Traditional & ANC(RLS) \& AS-LMS \cite{b5ram2011novel} &
Reference-based adaptive cancellation (RLS); variable step-size LMS. &
PPG$+$ACC &
MA-reduced PPG; HR/SpO$_2$ &
General activity \\

Traditional & EPDA \cite{b36bradley2024opening} &
Efficient envelope-based PPG denoising algorithm. &
PPG &
MA-reduced PPG; HR &
N/A \\

Traditional & TROIKA \cite{b6zhang2014troika} &
Sparse spectrum reconstruction$+$spectral peak tracking with ACC assistance. &
PPG$+$ACC &
HR &
Low--high \\

Traditional & JOSS \cite{b25zhang2015photoplethysmography} &
Jointly estimates PPG and accelerometer spectra via MMV-based sparse recovery. &
PPG$+$ACC &
HR &
Low--high \\

Traditional & CEEMDAN \cite{b35zhang2024removal} &
CEEMDAN-based entropy feature and adaptive filter for removal. &
PPG$+$ACC &
Improved SpO$_2$ accuracy; MA-reduced PPG &
Everyday activity \\

Traditional & ANFA \cite{b18zheng2022adaptive} &
Adaptive frequency-domain MA removal with ACC assistance. &
PPG$+$ACC &
MA-free PPG; HR/RR &
Low--high \\

Traditional & MR \cite{b21zheng2023rapid} &
Suppresses MAs by using mOEPS and ACC estimates to select a pulsatile component with minimal residuals. &
PPG$+$ACC $+$VEL &
MA-free PPG; HR/RR/SpO$_2$ &
Low--high \\
\midrule
\multicolumn{6}{l}{\textit{Machine-learning / Deep-learning approaches}}\\
\midrule
ML/DL (End-to-end) & End-to-end mapping \cite{b8luque2018end} &
DL regression from raw PPG to target vital signs. &
PPG &
HR/RR/SpO$_2$ &
N/A \\

ML/DL (Features) & Feature-extraction DL \cite{b9biswas2019cornet} &
CNN-based feature learning from pre-processed PPG. &
PPG &
HR/RR/SpO$_2$ &
N/A \\

DL (End-to-end) & DeepPPG \cite{b29reiss2019deep} &
End-to-end model uses synchronized PPG/ACC spectrograms to estimate HR. &
PPG$+$ACC &
HR &
N/A \\

DL (Quality) & Robust 1D-CNN \cite{b11goh2020robust} &
1D-CNN signal-quality classifier (clean vs artefact). &
PPG &
Quality label; HR &
Low \\

DL (Quality, 2D) & 2D representations \cite{b12chen2021signal} &
2D transform $+$ 2D-CNN for quality at rest/low intensity. &
2D-PPG &
Quality label; HR &
Rest--low \\

GAN (denoising) & Corr-image GAN \cite{b13afandizadeh2023accurate} &
Convert PPG to 2D correlation images; image GAN for MA removal. &
2D-PPG &
MA-reduced PPG; HR &
Low \\

GAN (PPG MA removal) & PPG-GAN \cite{b29zheng2022ppg} &
GAN to generate de-noised PPG without ACC/Gyro. &
PPG &
MA-reduced PPG; HR/RR &
Low--high \\

Attention (HR) & PULSE \cite{b37kasnesis2023feature} &
Temporal convolutions $+$ multi-head attention for HR. &
PPG &
HR &
N/A \\

Attention (HR) & Adaptive Q-PPG \cite{b38kechris2024kid} &
Attention-enhanced Q-PPG for HR tracking. &
PPG &
HR &
N/A \\

GAN$+$Attention (synthesis) & CardioGAN \cite{b31sarkar2021cardiogan} &
Attention generator; dual discriminators (time$+$freq) for ECG-from-PPG synthesis. &
PPG &
Synthesised ECG &
Rest \\

GAN$+$Attention (EEG) & EEGANet \cite{b17sawangjai2021eeganet} &
GAN-based ocular artefact removal (EEG). &
EEG &
MA-reduced EEG &
N/A \\

GAN$+$Attention (PPG) & Attention GAN \cite{b33sawangjai2025removal} &
Attention GAN with dual discriminator for MA removal without ACC/Gyro. &
PPG &
MA-reduced PPG; HR &
Low--medium \\
\midrule
\multicolumn{6}{l}{\textit{This work (proposed)}}\\
\midrule
GAN$+$Attention (fusion) & AM-GAN &
Attention within GAN for multi-sensor fusion and MA removal. &
PPG$+$ACC $+$VEL &
MA-free PPG; HR/RR/SpO$_2$ &
Low--high \\
\bottomrule
\end{tabular}
\end{adjustbox}
\end{table}

\clearpage
\bibliographystyle{elsarticle-num} 
\bibliography{main}





\end{document}